\documentclass{ws-mod-ehs}

\newcommand{\beq}{\begin{equation}}
\newcommand{\eeq}{\end{equation}}
\newcommand{\beqa}{\begin{eqnarray}}
\newcommand{\eeqa}{\end{eqnarray}}

\newcommand{\laem}{\begin{array}{c} < \vspace{-0.5em} \\ {\scriptstyle \sim}
\end{array}}
\newcommand{\gaem}{\begin{array}{c} > \vspace{-0.5em} \\ {\scriptstyle \sim}
\end{array}}

\def\slashchar#1{\setbox0=\hbox{$#1$}           
   \dimen0=\wd0                                 
   \setbox1=\hbox{/} \dimen1=\wd1               
   \ifdim\dimen0>\dimen1                        
      \rlap{\hbox to \dimen0{\hfil/\hfil}}      
      #1                                        
   \else                                        
      \rlap{\hbox to \dimen1{\hfil$#1$\hfil}}   
      /                                         
   \fi}                                         %

\newcommand{\tr}{{\rm Tr}}

\newcommand{\mev}{{\rm \, MeV}}
\newcommand{\gev}{{\rm \, GeV}}
\newcommand{\tev}{{\rm \, TeV}}
\newcommand{\W}{{\rm\bf W}}

\def\be{\begin{equation}}
\def\ee{\end{equation}}
\def\bea{\begin{eqnarray}}
\def\eea{\end{eqnarray}}

\begin{document}

\title{Lectures on Technicolor and Compositeness}

\author{R. Sekhar Chivukula}

\address{Physics Department\\ 
Boston University\\
590 Commonwealth Ave.\\
Boston, MA 02215, USA\\
E-mail: sekhar@bu.edu\\
{\tt BUHEP-00-24}}


\maketitle

\abstracts{
Lecture 1 provides an introduction to dynamical
electroweak symmetry breaking. Lectures 2 and 3
give an introduction to compositeness, with emphasis on 
effective lagrangians, power-counting, and the
't Hooft anomaly-matching conditions.
}

\bigskip
\noindent{\bf \underline{Lecture 1: Technicolor}\footnote{What follows is largely an
    abbreviated version of the sections on technicolor in
    lectures\cite{Chivukula:1998if} I presented at the Les Houches
    summer school in 1997. For lack of space, I have not included a
    description of the phenomenology of dynamical electroweak symmetry
    breaking -- for a recent review see Chivukula and Womersley in the
    2000 Review of Particle Properties.\cite{Groom:2000in}}}

\section{Dynamical Electroweak Symmetry Breaking}

The simplest theory of dynamical electroweak symmetry
breaking is technicolor.\cite{Weinberg:1979bn,Susskind:1978ms} Consider
an $SU(N_{TC})$ gauge theory with fermions in the fundamental
representation of the gauge group
\beq
\Psi_L=\left(
\begin{array}{c}
U\\D
\end{array}
\right) _L\,\,\,\,\,\,\,\,
U_R,D_R~.
\eeq
The fermion kinetic energy terms
for this theory are
\beqa
{\cal L} &=& \bar{U}_L i\slashchar{D} U_L+
\bar{U}_R i\slashchar{D} U_R+\\
 & &\bar{D}_L i\slashchar{D} D_L+
\bar{D}_R i\slashchar{D} D_R~,
\nonumber
\eeqa
and, like QCD in the $m_u$, $m_d \to 0$ limit, they have
a chiral $SU(2)_L \times SU(2)_R$ symmetry.

As in QCD, exchange of technigluons in the spin zero, isospin zero
channel is attractive, causing the formation of a condensate
\beq
{\lower15pt\hbox{
\epsfysize=0.5 truein \epsfbox{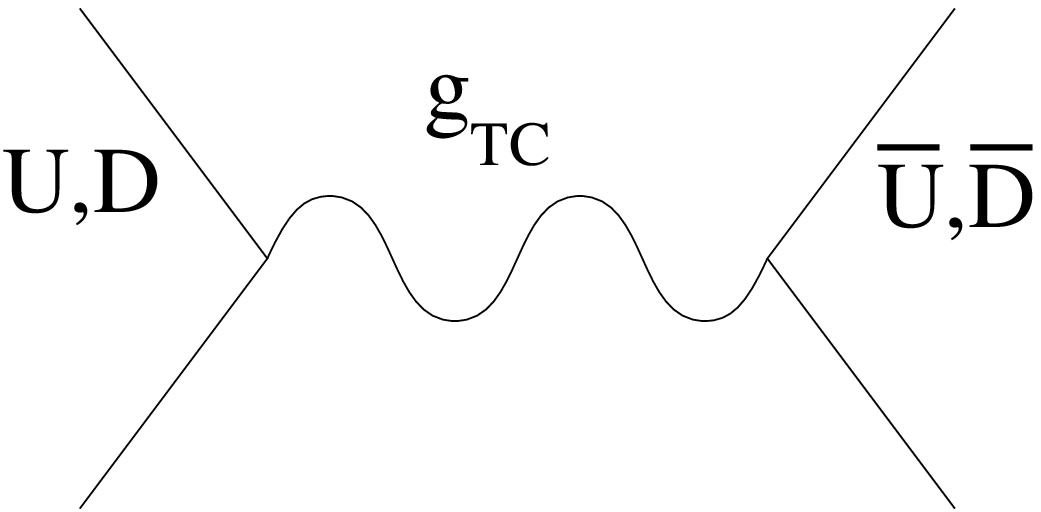}}
\ \ \rightarrow  \ \ \langle \bar U_LU_R\rangle
=\langle \bar D_LD_R\rangle \neq 0\, ,}
\eeq
which dynamically breaks $SU(2)_L \times SU(2)_R \to SU(2)_V$.  These
broken chiral symmetries imply the existence of three massless Goldstone
bosons, the analogs of the pions in QCD.

Now consider gauging $SU(2)_W \times U(1)_Y$ with the left-handed
fermions transforming as weak doublets and the right-handed ones as weak
singlets. To avoid gauge anomalies, in this one-doublet technicolor
model we will take the left-handed technifermions to have hypercharge
zero and the right-handed up- and down-technifermions to have
hypercharge $\pm 1/2$.  The spontaneous breaking of the chiral symmetry
breaks the weak-interactions down to electromagnetism. The would-be
Goldstone bosons become the longitudinal components of the $W$ and $Z$
\beq
\pi^\pm,\, \pi^0 \, \rightarrow\, W^\pm_L,\, Z_L~,
\eeq
which
acquire a mass
\beq
M_W = {g F_{TC} \over 2}~.
\eeq
Here $F_{TC}$ is the analog of $f_\pi$ in QCD. In order
to obtain the experimentally observed masses, we must have
$F_{TC} \approx 246 {\rm GeV}$ and hence this model
is essentially QCD scaled up by a factor of
\beq
{F_{TC}\over f_\pi} \approx 2500\, .
\eeq

While I have described only the simplest model above, it is straightforward
to generalize to other cases.  {\it Any} strongly interacting gauge theory
with a chiral symmetry breaking pattern $G \to H$, in which $G$ contains
$SU(2)_W \times U(1)_Y $ and subgroup $H$ contains $U(1)_{em}$ (but not
$SU(2)_W \times U(1)_Y$) will break the weak interactions down to
electromagnetism.  In order to be consistent with experimental results,
however, we must also require that $H$ contain
custodial\cite{Weinstein:1973gj,Sikivie:1980hm} $SU(2)_V$. This custodial
symmetry insures that the $F$-constant associated with the $W^\pm$ and $Z$
are equal and therefore that the relation
\beq
\rho = {M^2_W \over M^2_Z \cos^2\theta_W} =1
\label{custodial}
\eeq
is satisfied at tree-level.  If the chiral symmetry is larger than
$SU(2)_L\times SU(2)_R$, theories of this sort will contain additional
(pseudo-) Goldstone bosons which are not ``eaten'' by the $W$ and $Z$.

\section{Flavor Symmetry Breaking and ETC}

\subsection{Fermion Masses \& ETC Interactions}

In order to give rise to masses for the ordinary quarks and leptons, we
must introduce interactions which connect the chiral-symmetries of
technifermions to those of the ordinary fermions. The most popular
choice\cite{Eichten:1979ah,Dimopoulos:1979es} is to introduce new
broken gauge interactions, called {\it extended technicolor
  interactions} (ETC), which couple technifermions to ordinary fermions.
At energies low compared to the ETC gauge-boson mass, $M_{ETC}$, these
effects can be treated as local four-fermion interactions
\beq
{\lower15pt\hbox{\epsfysize=0.5 truein \epsfbox{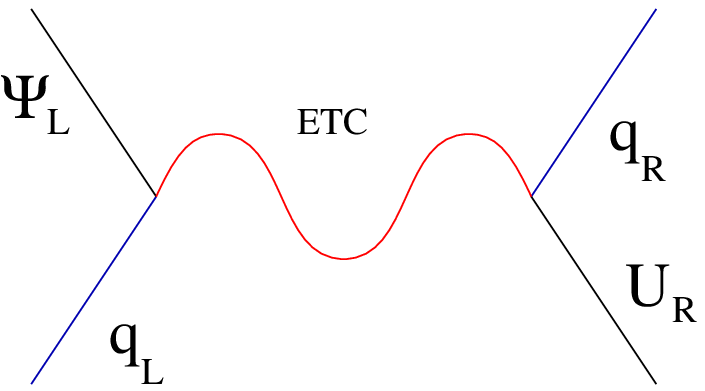}}}
\ \ \rightarrow\ \  {{g_{ETC}^2\over M^2_{ETC}}}(\overline{\Psi}_L U_R)
({\overline{q}_R q_L})~.
\label{etcint}
\eeq
After technicolor chiral-symmetry breaking and the formation of a
$\langle \bar{U} U \rangle$ condensate, such an interaction gives rise
to a mass for an ordinary fermion
\beq
m_q \approx {{g_{ETC}^2\over M^2_{ETC}}} \langle\overline{U} U\rangle_{ETC}~,
\label{fmass}
\eeq
where $\langle \overline{U} U\rangle_{ETC}$ is the value of the
technifermion condensate evaluated at the ETC scale (of order
$M_{ETC}$).  The condensate renormalized at the ETC scale in eq.
(\ref{fmass}) can be related to the condensate renormalized at the
technicolor scale as follows
\beq
\langle\overline{U} U\rangle_{ETC} = \langle\overline{U} U\rangle_{TC}
\exp\left(\int_{\Lambda_{TC}}^{M_{ETC}} {d\mu \over \mu}
\gamma_m(\mu)\right)~,
\eeq
where $\gamma_m(\mu)$ is the anomalous dimension of the
fermion mass operator and $\Lambda_{TC}$ is the analog of $\Lambda_{QCD}$
for the technicolor interactions.

For QCD-like technicolor (or any theory which is ``precociously''
asymptotically free), $\gamma_m$ is small  in the range between
$\Lambda_{TC}$ and $M_{ETC}$. Using dimensional analysis
\cite{Weinberg:1979kz,Georgi:1985kw,Manohar:1984md,Gasser:1985gg}
we find
\beq
\langle\overline{U} U\rangle_{ETC} \approx \langle\overline{U} U\rangle_{TC}
\approx 4\pi F^3_{TC}~.
\eeq
In this case eq. (\ref{fmass}) implies that
\beq
{{M_{ETC}\over g_{ETC}}} \approx 40 \tev 
\left({F_{TC}\over 250\gev}\right)^{3\over 2}
\left({100 \mev \over m_q}\right)^{1\over 2}~.
\eeq

In order to orient our thinking, it is instructive to consider a simple
``toy'' extended technicolor model. The model is based on an $SU(N_{ETC})$
gauge group, with technicolor as an extension of flavor.  In this case
$N_{ETC} = N_{TC} + N_F$, and the model contains the (anomaly-free) set of
fermions 
\beq
\begin{array}{l@{\extracolsep{15pt}}l}
Q_L=(N_{ETC},3,2)_{1/6} & L_L=(N_{ETC},1,2)_{-1/2} \\
U_R=(N_{ETC},3,1)_{2/3} & E_R=(N_{ETC},1,1)_{-1} \\
D_R=(N_{ETC},3,1)_{-1/3} & N_R=(N_{ETC},1,1)_{0}~,
\end{array}
\label{onefamily}
\eeq
where we display their quantum numbers under $SU(N_{ETC})\times
SU(3)_C \times SU(2)_W \times U(1)_Y$. We break the
ETC group down to technicolor in three stages
\medskip
\begin{center}
{$SU(N_{TC}+3)$}
\end{center}
\begin{center}
$\Lambda_1\ \ \ \ \ \downarrow \ \ \ \ \  
m_1\approx{4\pi F^3\over \Lambda^2_1}$
\end{center}
\begin{center}
{$SU(N_{TC}+2)$}
\end{center}
\begin{center}
$\Lambda_2\ \ \ \ \ \downarrow \ \ \ \ \  
m_2\approx{4\pi F^3\over \Lambda^2_2}$
\end{center}
\begin{center}
{$SU(N_{TC}+1)$}
\end{center}
\begin{center}
$\Lambda_3\ \ \ \ \ \downarrow \ \ \ \ \  
m_3\approx{4\pi F^3\over \Lambda^2_3}$
\end{center}
\begin{center}
{$SU(N_{TC})$}
\end{center}
\medskip\noindent
resulting in three isospin-symmetric families of degenerate
quarks and leptons, with $m_1 < m_2 < m_3$. Note that the
{\it heaviest} family is related to the {\it lightest} ETC 
scale!

Before continuing our general discussion, it is worth noting
a couple of points. First,
in this example the ETC gauge bosons do not carry color
or weak charge
\beq
[G_{ETC},SU(3)_C]=[G_{ETC},SU(2)_W]=0~.
\label{commute}
\eeq
Furthermore, in this model there is one technifermion for each type of
ordinary fermion: that is, this is a ``one-family'' technicolor
model.\cite{Farhi:1979zx} Since there are eight left- and right-
handed technifermions, the chiral symmetry of the technicolor theory
is (in the limit of zero QCD and weak couplings) $SU(8)_L \times
SU(8)_R \to SU(8)_V$. Such a theory would yield $8^2-1=63$ (pseudo-)
Goldstone bosons. Three of these Goldstone bosons are unphysical ---
the corresponding degrees of freedom become the longitudinal
components of the $W^\pm$ and $Z$ by the Higgs mechanism.  The
remaining 60 must somehow obtain a mass. This will lead to the
condition in eq.  (\ref{commute}) being modified in a realistic
model.\cite{Eichten:1979ah} We will return to the issue of
pseudo-Goldstone bosons below.

The most important feature of this or any ETC-model is that a successful
extended technicolor model will provide a {\it dynamical theory of flavor}!
As in the toy model described above  and as explicitly
shown in eq. (\ref{etcint}) above, the masses of the ordinary fermions are
related to the masses and couplings of the ETC gauge-bosons. A successful and
complete ETC theory would predict these quantities and, hence, the ordinary
fermion masses.

Needless to say, constructing such a theory is very difficult. No
complete and successful theory has been proposed.  Examining our toy
model, we immediately see a number of shortcomings of this model that
will have to be addressed in a more realistic theory:

\begin{narrower}
\begin{itemize}
\item What breaks ETC?
\item Do we require a {separate} scale for each family?
\item How do the $T_3 = \pm {1\over 2}$ fermions of a given generation
receive {\it different} masses?
\item How do we obtain quark mixing angles?
\item What about right-handed technineutrinos and $m_\nu$?
\end{itemize}
\end{narrower}

\subsection{Flavor-Changing Neutral-Currents}

Perhaps the single biggest obstacle to constructing a realistic ETC
model (or any dynamical theory of flavor) is the potential for
flavor-changing neutral currents.\cite{Eichten:1979ah}  Quark mixing implies
transitions between different generations: $q \to \Psi \to q^\prime$,
where $q$ and $q'$ are quarks of the same charge from different
generations and $\Psi$ is a technifermion. Consider the commutator of
two ETC gauge currents:
\beq
[\overline{q}\gamma \Psi, \overline{\Psi}\gamma q^\prime]\, \supset\, 
\overline{q}\gamma q^\prime\, .
\eeq
Hence we expect there are ETC gauge bosons which couple to flavor-changing
neutral currents. In fact, this argument is slightly too slick: the same
applies to the charged-current weak interactions!  However in that case the
gauge interactions, $SU(2)_W$, respect a global $(SU(3) \times U(1))^5$ chiral
symmetry\footnote{One $SU(3)$ flavor symmetry for the three families of each
  type of ordinary fermion.\protect\cite{Chivukula:1987py}} leading to the
usual {GIM} mechanism.

Unfortunately, the ETC interactions {cannot} respect the same global symmetry;
they must distinguish between the various generations in order to give
rise to the masses of the different generations. Therefore, flavor-changing
neutral-current interactions are (at least at some level) unavoidable.

The most severe constraints come from possible $|\Delta S| = 2$
interactions which contribute to the $K_L$-$K_S$ mass
difference. In particular, we would expect that in order to produce
Cabibbo-mixing the same interactions which give rise to the $s$-quark
mass could cause the flavor-changing interaction
\beq
{\cal L}_{\vert \Delta S \vert = 2} = {g^2_{ETC} \, \theta^2_{sd} \over
{M^2_{ETC}}} \,\, \left(\overline{s} \Gamma^\mu d\right) \,\, 
\left(\overline{s} \Gamma'_\mu d\right) + {\rm
h.c.}~,
\eeq
where $\theta_{sd}$ is of order the Cabibbo angle. Such an interaction
contributes to the neutral kaon mass splitting
\beq 
(\Delta M^2_K)_{ETC} =  {g^2_{ETC} \, \theta^2_{sd} \over
{M^2_{ETC}}} \, \langle \overline{K^0} \vert \overline{s} \Gamma^\mu d \, \overline{s}
\Gamma'_\mu d \vert K^0 \rangle + {\rm c.c.}
\eeq
Using the vacuum insertion approximation we find
\beq
(\Delta M^2_K)_{ETC} \simeq {g^2_{ETC} \, 
{\rm Re}(\theta^2_{sd}) \over {2 M^2_{ETC}}} \,
f^2_K M^2_K ~.
\eeq
Experimentally\cite{Groom:2000in} we know that
$\Delta M_K <   3.5 \times 10^{-12}\,\mev $ and, hence, that
\beq
{M_{ETC} \over {g_{ETC} \, \sqrt{{\rm Re}(\theta^2_{sd})}}} >  600\,\tev
\eeq
Using eq. (\ref{fmass}) we find that
\beq
m_{q, \ell} \simeq {g_{ETC}^2 \over {M_{ETC}^2}}
\langle\overline{T}T\rangle_{ETC}  < {0.5\,\mev\over{N_D^{3/2} \, \theta_{sd}^2}} \,
\eeq
showing that it will be difficult to produce the {$s$}-quark
mass, let alone the {$c$}-quark!

\subsection{Pseudo-Goldstone Bosons}

A ``realistic'' ETC theory may require a technicolor sector with a chiral
symmetry structure bigger than the $SU(2)_L \times SU(2)_R$ discussed
initially. The prototypical theory has one-family of technifermions, as
incorporated in our toy model. As discussed there, the theory has an $SU(8)_L
\times SU(8)_R \to SU(8)_V$ chiral symmetry breaking structure resulting in
63 Goldstone bosons, 3 of which are unphysical. The quantum numbers of the 60
remaining Goldstone bosons are shown in table 1. Clearly, these
objects cannot be massless in a realistic theory!

\begin{table}
\begin{center}
\begin{tabular}{@{}lll@{}}   \hline
SU$(3)_C$   & SU$(2)_{V}$ &Particle         \\ \hline
$1$      &$1$  &$P^{0 \prime} \;,\; \omega_T$  \\
$1$      &$3$  &$P^{0,\pm} \;,\; \rho^{0,\pm}_T$  \\
$3$      &$1$  &$P^{0 \prime}_{3} \;,\; \rho^{0 \prime }_{T 3}$  \\
$3$      &$3$  &$P^{0,\pm}_{3} \;,\; \rho^{0,\pm}_{T 3}$  \\
$8$      &$1$  &$P^{0 \prime}_{8} (\eta_T) \;,\; \rho^{0 \prime }_{T 8}$  \\
$8$      &$3$    &$P^{0,\pm}_{8} \;,\; \rho^{0,\pm}_{T 8}$  \\ \hline
\end{tabular}
\label{pgbtab}
\caption{Quantum numbers of the 60 physical Goldstone bosons (and the
corresponding vector mesons) in a one-family technicolor model. Note
that the mesons that transform as  3's of QCD are complex fields.}
\end{center}
\end{table}

In fact, the ordinary gauge interactions break the full $SU(8)_L
\times SU(8)_R$ chiral symmetry explicitly. The largest effects are
due to QCD, and the color octets and triplets mesons get masses of
order 200 -- 300 GeV, in analogy to the electromagnetic mass splitting
$m_{\pi^+}-m_{\pi^0}$ in QCD. Unfortunately, the
others\cite{Eichten:1979ah} are {massless} to {\cal O}($\alpha$)!

Luckily, the ETC interactions (which we introduced in order to give
masses to the ordinary fermions) are capable of explicitly breaking
the unwanted chiral symmetries and producing masses for these mesons.
This is because in addition to coupling technifermions to ordinary
fermions, some ETC interactions also couple technifermions to
one another.\cite{Eichten:1979ah} Using Dashen's
formula,\cite{Dashen:1969eg} we can estimate that such an interaction
can give rise to an effect of order
\beq
F^2_{TC} M^2_{\pi_T} \propto {g^2_{ETC} \over M^2_{ETC}}
\langle (\overline{T}T)^2\rangle_{ETC}~.
\label{dashen}
\eeq
In the vacuum insertion approximation for a theory with small
$\gamma_m$, we may rewrite the above formula using eq. (\ref{fmass}) and
find that
\beq
M_{\pi_T} \simeq 55\gev 
\sqrt{m_f \over 1\gev} \sqrt{250 \gev \over F_{TC}}~.
\label{tpimass}
\eeq
It is unclear whether this is large enough.

In addition, there is a particularly troubling chiral symmetry
in the one-family model. The 
$SU(8)$-current $\overline{Q}\gamma_\mu \gamma_5 Q - 3 \overline{L} \gamma_\mu
\gamma_5 L$ is {spontaneously broken} and {has a color
  anomaly}. Therefore, we have a potentially
dangerous {weak scale axion\cite{Peccei:1977hh,Peccei:1977ur,Weinberg:1978ma,Wilczek:1978pj}}!
An ETC-interaction of the form
\beq
{g^2_{ETC} \over M^2_{ETC}} \left(\overline{Q}_L \gamma^\mu L_L \right)
\left(\overline{L}_R \gamma^\mu Q_R \right)~,
\eeq
is required to give to an axion mass, and
{we must\cite{Eichten:1979ah} embed $SU(3)_C$ in $ETC$.}

\subsection{ETC etc.}

There are other model-building constraints\cite{Lane:1993wz} on a
realistic TC/ETC theory. A realistic ETC theory:

\begin{narrower}
\begin{itemize}

\item must be asymptotically free,

\item cannot have gauge anomalies,

\item must produce small neutrino masses, 

\item cannot give rise to extra massless (or even light) gauge bosons,

\item should generate weak CP-violation without producing unacceptably
  large amounts of strong CP-violation,

\item must give rise to isospin-violation in fermion masses without
  large contributions\cite{Appelquist:1984nc,Appelquist:1985rr} to $\Delta\rho$ and,
  
\item must accommodate a large $m_t$ while giving rise to only small
  corrections\cite{Chivukula:1992ap,Randall:1993ua} to $Z\to \overline{b}b$
  and $b\to s\gamma$.

\end{itemize}
\end{narrower}

Clearly, building a fully realistic ETC model will be quite difficult!
However, as I have emphasized before, this is because an ETC theory must
provide a complete dynamical explanation of flavor.  In the next
section, I will concentrate on possible solutions to the flavor-changing
neutral-current problem(s). 

\section{Walking Technicolor}

\subsection{The Gap Equation}

Up to now we have assumed that technicolor is, like QCD, precociously
asymptotically free with a small anomalous dimension $\gamma_m(\mu)$ for
scales $\Lambda_{TC} < \mu < M_{ETC}$. However, as discussed above it is
difficult to construct an ETC theory of this sort without producing
dangerously large flavor-changing neutral currents. On the other hand, if the
$\beta$-function $\beta_{TC}$ is {\it small}, $\alpha_{TC}$ can remain large
above the scale $\Lambda_{TC}$ --- {\it i.e.} the technicolor coupling would
``walk'' instead of running. In this same range of momenta, $\gamma_m$ may be
large and, since 
\beq 
\langle\overline{T} T\rangle_{ETC} =
\langle\overline{T} T\rangle_{TC}\, \exp\left(\int_{\Lambda_{TC}}^{M_{ETC}}
  {d\mu \over \mu} \gamma_m(\mu)\right) 
\eeq
this could enhance the size of the condensate renormalized at the ETC scale
($\langle \overline{T} T\rangle_{ETC}$) and produce larger fermion
masses.\cite{Holdom:1981rm,Holdom:1985sk,Yamawaki:1986zg,Appelquist:1986an,Appelquist:1987tr,Appelquist:1987fc}

\begin{figure}[tbp]
\hskip80pt\epsfxsize=6cm\epsfbox{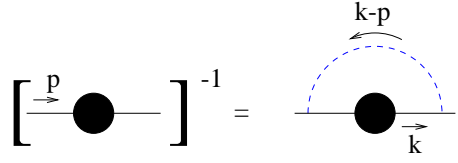}
\caption{Schwinger-Dyson equation for the
fermion self-energy function $\Sigma(p)$ in the rainbow
approximation. The dashed line represents the technigluon propagator and
the solid line technifermion propagator.}
\label{Fig11}
\end{figure}

In order to proceed further, however, we need to understand how large
$\gamma_m$ can be and how walking affects the technicolor chiral symmetry
breaking dynamics.  These questions cannot be addressed in perturbation
theory.  Instead, what is conventionally done is to use a 
nonperturbative approximation for $\gamma_m$ and chiral-symmetry breaking
dynamics based on the ``rainbow'' approximation\cite{Pagels:1975se,Peskin:1982mu}
to the Schwinger-Dyson equation shown in Figure \ref{Fig11}.
Here we write the full, nonperturbative, 
fermion propagator in momentum space as
\beq
iS^{-1}(p) = Z(p)(\slashchar{p} - \Sigma(p))~.
\eeq
The linearized form of the gap equation in
Landau gauge (in which $Z(p) \equiv 1$ in the rainbow
approximation) is
\beq
\Sigma(p) = 3 C_2(R)\, \int {d^4 k \over {(2 \pi)^4}}
\, {\alpha_{TC}((k-p)^2) \over {(k-p)^2}} \, {\Sigma(k) \over {k^2}}~.
\eeq
Being separable, this integral equation can be converted to a
differential equation which has the approximate (WKB)
solutions\cite{Fukuda:1976zb,Higashijima:1984gx}
\beq
\Sigma(p) \propto p^{-\gamma_m(\mu)}\, ,\, \, p^{\gamma_m(\mu)-2}\, .
\label{twosol}
\eeq
Here $\alpha(\mu)$ is assumed to run slowly, as
will be the case in walking technicolor, and
the anomalous dimension of the fermion mass operator is
\beq
\gamma_m(\mu)=1-\sqrt{1-{\alpha_{TC}(\mu)\over\alpha_C}}\, ; \ \ \ \ \ \alpha_C
\equiv {\pi \over 3 C_2(R)}\, .
\label{crit}
\eeq

One can give a physical interpretation of these two
solutions\cite{Lane:1974he,Politzer:1976tv} in eq. \ref{twosol}. Using the
operator product expansion, we find
\beq
\lim_{p\to\infty} \Sigma(p) \ \ \propto \ \ 
\lower15pt\hbox{\epsfxsize=4cm\epsfbox{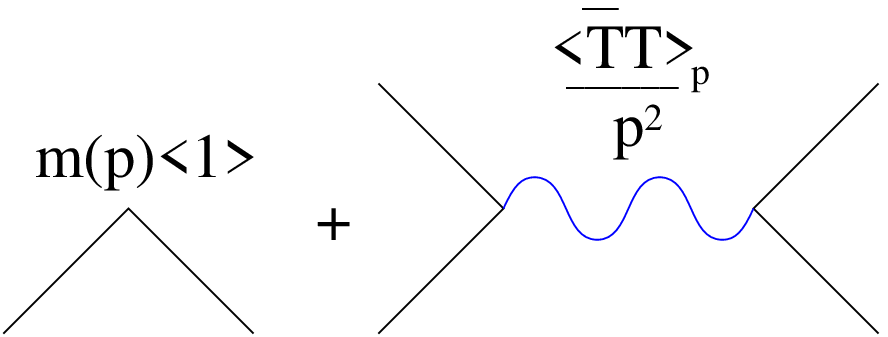}}~.
\eeq
Thus the first solution corresponds to a ``hard mass'' or explicit
chiral symmetry breaking, while the second solution corresponds to a
``soft mass'' or spontaneous chiral symmetry breaking.  If we let $m_0$
be the explicit mass of a fermion, dynamical symmetry breaking occurs
only if
\beq
\lim_{m_0 \to 0} \Sigma(p) \neq 0\, .
\eeq
A careful analysis of the gap equation, or equivalently the appropriate
effective potential,\cite{Cornwall:1974vz} implies that this happens only if
$\alpha_{TC}$ reaches the critical value of chiral
symmetry breaking, $\alpha_C$ defined in eq. (\ref{crit}).
Furthermore, the chiral symmetry breaking scale $\Lambda_{TC}$ is
defined by the scale at which
\beq
\alpha_{TC}(\Lambda_{TC})=\alpha_C 
\eeq
and, hence, at least in the rainbow approximation, at which
\beq
\gamma_m(\Lambda_{TC})=1.
\eeq
In the rainbow approximation, then, chiral symmetry breaking occurs when the
{``hard''} and {``soft''} masses scale the same way. It is believed that even
beyond the rainbow approximation one will find $\gamma_m =1$ at the critical
coupling.\cite{Appelquist:1988yc,Cohen:1989sq,Mahanta:1989rb}

\subsection{Implications of Walking: Fermion and PGB Masses}

If $\beta(\alpha_{TC}) \simeq 0$ all the way from $\Lambda_{TC}$
to $M_{ETC}$, then  $\gamma_m(\mu) \cong 1$ in this
range. In this case, eq. (\ref{fmass}) becomes
\beq
m_{q,l} = {g^2_{ETC} \over {M^2_{ETC}}} \times
\left(\langle\overline{T}T\rangle_{ETC} \cong 
\langle\overline{T}T\rangle_{TC} \, {M_{ETC} \over {\Lambda_{TC}}} \right)~.
\eeq
We have previously estimated that flavor-changing
neutral current requirements imply that the
ETC scale associated with the second generation must
be greater than of order 100 to 1000 TeV. In the case of walking
technicolor the enhancement of the technifermion condensate implies that
\beq
m_{q,l} \simeq {{50\, -\, 500\mev}\over N^{3/2}_D \theta^2_{sd}}~,
\eeq
arguably enough to accommodate the strange and charm quarks.

In addition to modifying our estimate of the relationship
between the ETC scale and ordinary fermion masses, walking
also influences the size of pseudo-Goldstone boson masses. 
In the case of walking, Dashen's formula for the
size of pseudo-Goldstone boson masses in the presence
of chiral symmetry breaking from ETC interactions, eq. (\ref{dashen}),
reads:
\beqa
F^2_{TC} M^2_{\pi_T} & \propto & {g^2_{ETC} \over M^2_{ETC}}
\langle \left(\overline{T}T\right)^2\rangle)_{ETC} \nonumber \\
&\approx& {g^2_{ETC} \over M^2_{ETC}} 
\left(\langle\overline{T}T\rangle_{ETC}\right)^2 \nonumber \\
&\simeq&{g^2_{ETC} \over M^2_{ETC}}
{M^2_{ETC}\over \Lambda^2_{TC}}
\left(\langle\overline{T}T\rangle_{TC}\right)^2\, .
\eeqa
Consistent with the rainbow approximation, we have used the vacuum-insertion 
to estimate the strong matrix element.
Therefore we find
\beqa
M_{\pi_T} & \simeq &  g_{ETC} 
\left({4\pi F^2_{TC} \over \Lambda_{TC}}\right) \nonumber\\
&\simeq &g_{ETC} \left({750\gev \over N_D}\right)
\left({1\tev\over \Lambda_{TC}}\right)~,
\eeqa
{\it i.e.} walking also enhances the size of pseudo-Goldstone
boson masses! As shown in the discussion of eq. \ref{tpimass},
such an enhancement is welcome.

While this is very encouraging, two {caveats} should be kept in mind.
First, the estimates given are for the limit of {``extreme walking''},
{\it i.e.} assuming that the technicolor coupling walks all the way
from the technicolor scale $\Lambda_{TC}$ to the relevant ETC scale
$M_{ETC}$. To produce a more complete analysis, ETC-exchange must be
incorporated into the gap-equation technology in order to estimate
ordinary fermion masses. Studies of this sort are encouraging; it
appears possible to accommodate the first and second generation masses
without necessarily having dangerously large flavor-changing neutral
currents.\cite{Holdom:1981rm,Holdom:1985sk,Yamawaki:1986zg,Appelquist:1986an,Appelquist:1987tr,Appelquist:1987fc}
The second issue, however, is what about the third generation quarks,
the {top} and {bottom}?  Because of the large top-quark mass, further
refinements\cite{Chivukula:1998if} or modifications will be necessary
to produce a viable theory of dynamical electroweak symmetry breaking.
This issue remains the outstanding obstacle\footnote{As noted by
  Lane,\protect\cite{Lane:1993wz} we cannot to apply precision
  electroweak
  tests\protect\cite{Peskin:1990zt,Peskin:1992sw,Golden:1991ig,Holdom:1990tc,Dobado:1991zh}
  to directly constrain theories of walking technicolor.} in ETC or
any theory of flavor. Various models, including top condensate, top
seesaw, and top-color assisted technicolor have been proposed; many
are discussed by Elizabeth Simmons in her lectures in this volume.

\vfill\eject
\noindent{\bf \underline{Lectures 2 \& 3: Compositeness}}

\section{What is Compositeness?}

The relevant phenomenological question is: can any of the observed
gauge bosons, the quarks and leptons, or the Higgs boson (if it
exists) be composite particles? As we shall quantify in this lecture,
the fact that the standard model works well implies that a successful
theory must be one in which

\begin{itemize}
  
\item { the short distance degrees of freedom are not the same as the
    long distance degrees of freedom}, and

\item { the masses of the composite states are {\it much less}
than the intrinsic scale of the dynamics $\Lambda$.}

\end{itemize}

In order to obtain {\it light} bound states, the binding energy must
be comparable to the intrinsic scale $\Lambda$ and, therefore, the
bound states must be { relativistic} and the theory must be {
  strongly-coupled}.  We will be discussing field theories in which
these conditions are satisfied. 

As the masses of the composite states are less than the intrinsic scale
($\Lambda$) of the underlying dynamics, there must be a consistent
effective field theory\footnote{General reviews of effective field
  theory have been written by Howard
  Georgi\protect\cite{Georgi:1993qn}, David
  Kaplan\protect\cite{Kaplan:1995uv}, and Antonio
  Pich.\protect\cite{Pich:1998xt}} valid for energies $E < \Lambda$ describing
dynamics in that energy range.  In general, we will not be able to
completely solve the strongly-interacting underlying dynamics to give
a complete description of the low-energy properties of the bound
states. However, we may estimate the types and sizes of
interactions\cite{Weinberg:1979kz,Manohar:1984md,Georgi:1993dw,Chivukula:1992nw}
based on the following principles:

\begin{itemize}
  
\item {\bf That which is not forbidden is required}: the effective
lagrangian will include all interactions consistent with space-time,
global, and gauge symmetries (and, in the case of supersymmetric
theories, considerations of analyticity).

\item {\bf No small dimensionless numbers}: the interaction
  coefficients must be consistent with dimensional analysis.

\end{itemize}

When $\Lambda \to \infty$, {\it i.e.} if there is a large hierarchy of
scales, the effective theory must reduce to a {\it renormalizable} theory up
to corrections suppressed by powers of $\Lambda$. From this point of view,
the fact that current experimental results are consistent with a
renormalizable theory (the standard one-doublet higgs model) only implies
that the scale $\Lambda$ must be larger (perhaps substantially larger) than
energy scales we have experimentally probed.

\subsection{Dimensional Analysis}

Dimensional analysis is the key we will use to extract bounds on the
scale of compositeness $\Lambda$ from the results of experiments.
Using dimensional analysis, we will estimate { sizes} of interactions
involving composite scalars ($\phi$), fermions ($\psi$), and vector
bosons ($V^\mu$).  Since the effective theory should have no small
dimensionless numbers, the sizes of these interactions are determined
by {\it two} parameters:

\begin{itemize}

\item $\Lambda$, the {scale of the underlying strong dynamics}, and

\item $g$, the size of typical coupling constants.

\end{itemize}

As we now show, the natural size of $g$ is $4\pi$.
Let us start with\cite{Cohen:1997rt} the Wilsonian effective action at scale
$\Lambda$:
\beq
S_\Lambda = { {{\Lambda^4}\over g^2}}\int d^4 x\,  {\cal L}\left(
  {{ g}\phi \over { \Lambda}},{{ g} \psi \over {
      \Lambda^{3/2}}}, {{ g} V \over { \Lambda}}, {\partial
    \over { \Lambda}}\right)
\eeq
Here the parameters $\Lambda$ and $g$ are introduced to get the
dimensions correct and to account for an extra coupling for each field
in an interaction, while the $\Lambda^4/g^2$ is present to correctly
normalize the kinetic energy, {\it e.g.}:
\beq
\partial^\mu \phi^\dagger \partial_\mu \phi
= { \Lambda^4}{ {1 \over  g^2}} 
\left({\partial^\mu \over { \Lambda}}\right)
\left({{ g}\phi^\dagger \over { \Lambda}}\right)
\left({\partial_\mu \over { \Lambda}}\right) 
\left({{ g}\phi \over { \Lambda}}\right)~.
\eeq
Consider\cite{Weinberg:1979kz} a process which receives contributions
from one operator at tree-level, and another at $L$-loop order.  Any
powers of $\Lambda$ must be the same for both contributions.  The {
  $1/g^2$} pre-factor in $S_\Lambda$ ``counts'' the number of loops
and therefore the ratio of the $L$-loop and tree-level contributions
is of order
\beq 
\left({g^2\over 16\pi^2}\right)^{L}
\eeq
where we have included one $1/16\pi^2$ for each loop from 4-D phase space.
Neither of the two extreme possibilities for $g$ is self-consistent:
\begin{itemize}
  
\item $g \ll 4\pi$: in this case, the theory would be {\it weakly}
  coupled, in contradiction with our general expectation that theories
  with composite particles are strongly coupled;

\item If $g \gg 4\pi$: here our prescription for the sizes of various
  couplings would not be stable under a small change in the cutoff
  scale $\Lambda$.

\end{itemize}
Hence, we expect ${ g ={\cal O}( 4\pi)}$ is the natural size for
couplings in our effective theory.

In the Wilsonian effective theory, one computes with a momentum-space
cutoff of order $\Lambda$. {\it All} operators consistent with
symmetry requirements then contribute at the same order in $\Lambda^2$
to each process, making it impractical for use in actual calculations.
Instead, we will use a dimensionless regulator and organize
computation in powers of $\Lambda^{-1}$.  Matching our estimates from
such a calculation with those from the Wilsonian approach implies that
the rules of dimensional analysis give us the sizes of interaction
coefficients defined using a dimensionless regulator {\it renormalized
  at a scale of order $\Lambda$}.
  
In constructing the effective theory, we must impose all space-time,
global, and gauge symmetries by hand. We may also incorporate any
external, weakly-coupled fields ({\it e.g.}  the photon $A^\mu$), by
including an appropriate suppression factor ({\it e.g.} one factor of
$e/g$ for every $A^\mu$).

\subsection{Example: The QCD Chiral Lagrangian}

As an example of the use of dimensional analysis in an effective
lagrangian, consider the chiral lagrangian in
QCD.\cite{Coleman:1969sm,Callan:1969sn,Georgi:1985kw} The 
approximate $SU(2)_L \times SU(2)_R$ chiral symmetry
of the QCD lagrangian for light quarks is spontaneously broken to isospin,
$SU(2)_V$, producing three (approximate) Goldstone Bosons $\pi^a$
which we identify with the ordinary pions. In terms of the matrix
$\tilde{\pi} = \pi^a \sigma^a/2$, where the $\sigma^a$ are the Pauli
matrices, we define
\beq
\Sigma = \exp\left({{ 2}i{ g}\tilde{\pi}
\over { \Lambda}}\right)~,
\eeq
which transforms as
\beq
\Sigma \to { L} \Sigma { R^\dagger}\ \ \ \  
{ L},{ R} \in SU(2)_{{ L},{ R}}~.
\label{chiral}
\eeq

If we write the lagrangian for $\Sigma$ in an expansion
in powers of momentum, the lowest order term invariant
under the symmetry of eq. \ref{chiral} is
\beq
\hskip-20pt
{ 1\over 4}{{ \Lambda^4}\over { g^2}}{\rm Tr}\left[
\left({\partial^\mu \over { \Lambda}}\right)
\Sigma
\left({\partial_\mu \over { \Lambda}}\right)
\Sigma^\dagger
\right] =
{f^2_\pi \over 4} {\rm Tr}\left(\partial^\mu \Sigma \partial_\mu 
\Sigma^\dagger \right)~,
\eeq
where we have identified ${ \Lambda}/{ g} \equiv f_\pi \approx 93$
MeV (from the chiral current) and canonically normalized the kinetic
energy of the pion fields. It is customary to denote the dimensional
scale $g f_\pi$ by $\Lambda_{\chi SB} \simeq 1\, {\rm GeV}$.  Higher
order chirally invariant terms are possible and applying the
dimensional rules we find, for example, a term with four powers
of momentum
\beq
{f^2_\pi\over \Lambda^2_{\chi SB}}
{\rm Tr}\left(\partial^\mu \Sigma \partial_\mu 
\Sigma^\dagger \right)^2~.
\eeq
From this we conclude that chiral perturbation theory ($\chi$PT) an expansion
in $p^2/\Lambda^2_{\chi SB}$.\cite{Weinberg:1979kz,Manohar:1984md}

In reality, chiral symmetry is not exact, as the bare quark mass terms violate the
symmetry:
\beq
{\cal L}_{QCD} = \bar{\psi} i\slashchar{D} \psi -\bar{\psi}_L M \psi_R
-\bar{\psi}_R M^\dagger \psi_L~.
\eeq
We can incorporate $M$ as an ``external field'' in the chiral
lagrangian.\cite{Gasser:1985gg} Consider first the symmetry properties
of the quark mass term: ${\cal L}_{QCD}$ is ``invariant'' under a
chiral transformation combined with the redefinition
\beq
M \to { L} M { R^\dagger}\ \ \ \  
{ L},{ R} \in SU(2)_{{ L},{ R}}~,
\eeq
therefore terms incorporating $M$ must also have this property (this
is, essentially, an implementation of a generalized Wigner-Eckart
theorem). The power-counting for $M$ can be established by considering
the ``natural size'' of the fermion mass
\beq
{{ \Lambda^4}\over { g^2}}
\left({{ g}\bar{\psi} \over { \Lambda^{3/2}}}\right)
\left({{ g}\psi \over { \Lambda^{3/2}}}\right)
\equiv
\Lambda_{\chi SB}\,\bar{\psi}\psi~,
\eeq
in the absence of chiral symmetry.  Therefore, the small
parameter $M/\Lambda_{\chi SB}$ is a measure of explicit chiral
symmetry breaking and the leading term in $M$ has the form
\beq
{{ \Lambda^4}\over { g^2}}
\left[{\rm Tr}\left({M \over { \Lambda}}\Sigma^\dagger\right)
+\, h.c.\right]~.
\eeq
This leads to the usual result $m^2_\pi \simeq \Lambda_{\chi SB}(m_u +
m_d)$.


\section{The Phenomenology of Compositeness}
  
Using the rules of dimensional analysis, we can investigate the
phenomenology of the compositeness of the observed fermions and gauge
bosons. To the extent that they appear fundamental, we establish lower
bounds on their scale of compositeness.

\subsection{Fermions}

We begin by considering the quarks and leptons. Compositeness
can be expected to produce several phenomenological effects:

\begin{enumerate}

\item{Form Factors:} If ordinary fermions are composite particles,
we expect their gauge interactions to have nontrivial form factors.
These form factors can be thought of, in analogy with ``vector meson
dominance'' for the pion form-factor in QCD, as arising from processes
such as:
\beq
\includegraphics[width=2.5cm]{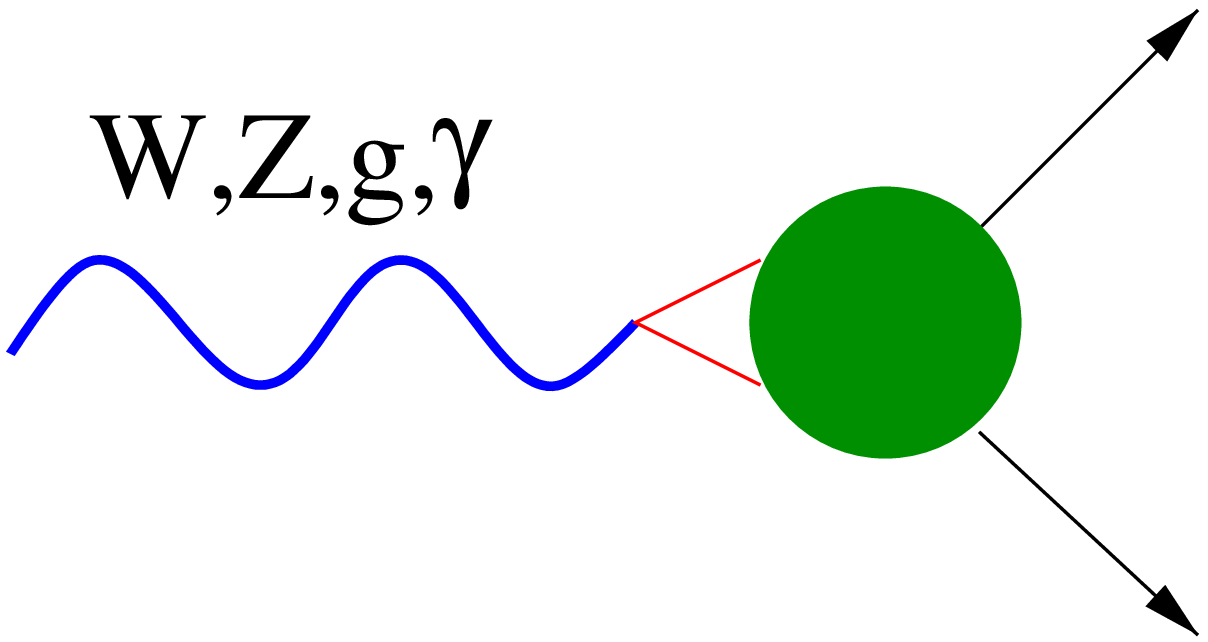}
\hskip+0.5cm
\raise+17pt\hbox{$\Rightarrow$}
\hskip+0.5cm
\raise+5pt\hbox{\includegraphics[width=3.5cm]{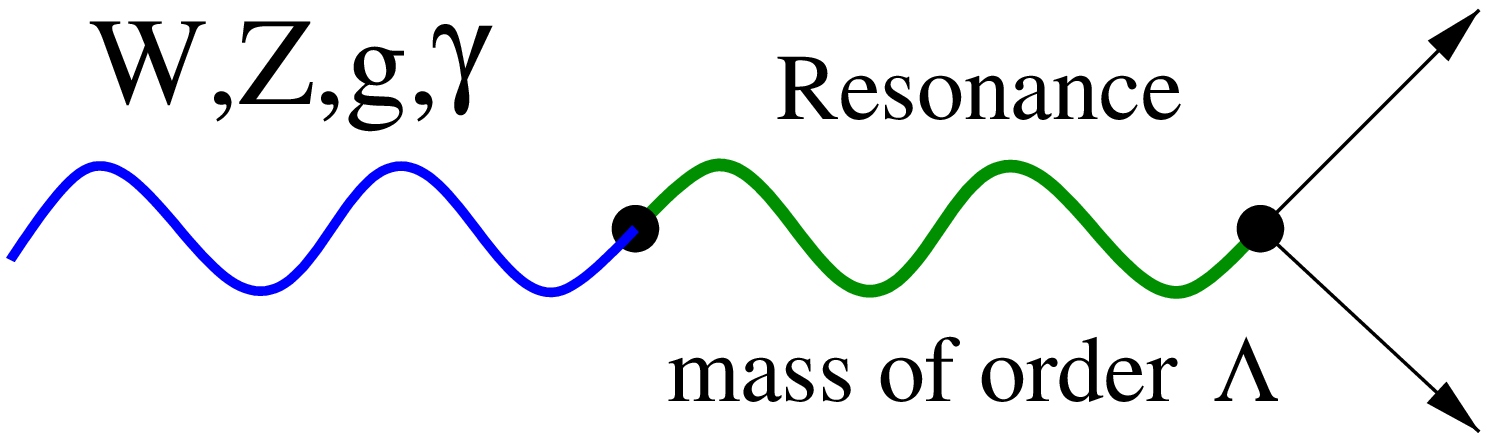}}
\eeq
yielding changes in four-fermion cross sections of the form
\beq
\hat{\sigma}(f\bar{f}\to f^\prime\bar{f^\prime}) \simeq \hat{\sigma}_{sm} 
\left[ 1 + {\cal O}\left({{{\hat{s}\over \Lambda^2}}} \right)\right]~,
\eeq
where $\hat{s}$ is the partonic center-of-mass energy squared of the
process.

\item{Contact Interactions\protect\cite{Eichten:1983hw}:} If ordinary
fermions are composite, they can also directly exchange heavy resonances
arising from the interactions responsible for binding the fermions:
\beq
\raise+2pt\hbox{\includegraphics[width=3.5cm]{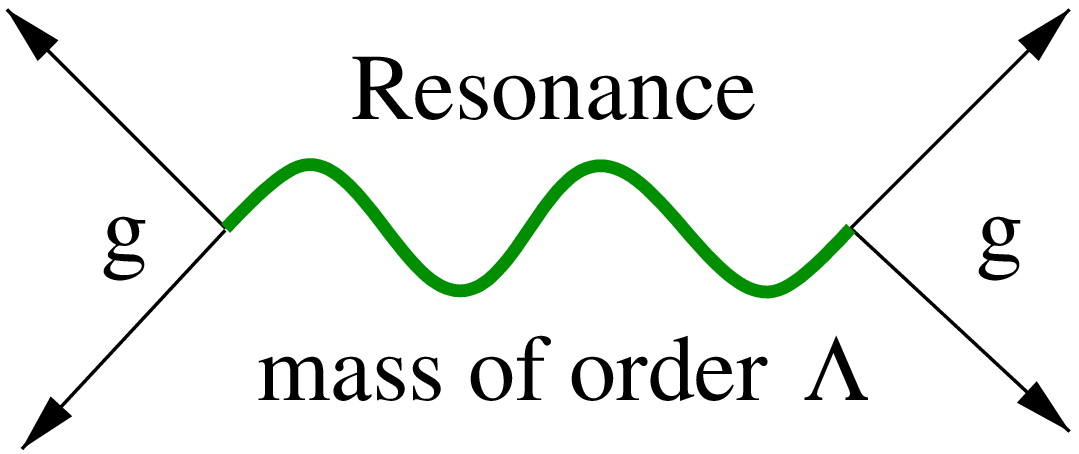}}
\hskip+0.5cm
\raise+20pt\hbox{$\Rightarrow$}
\hskip+0.5cm
\raise+0pt\hbox{\includegraphics[width=1.75cm]{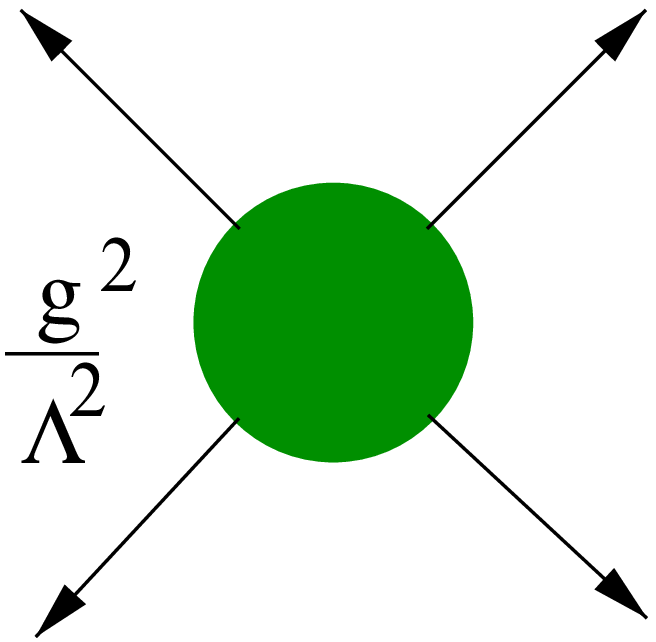}}~.
\eeq
The effect of these interactions on four fermion cross sections
\beq
\hat{\sigma}(f\bar{f}\to f^\prime\bar{f^\prime}) \simeq \hat{\sigma}_{sm} 
\left[ 1 + {\cal O}\left({{g^2\,\hat{s}\over 4\pi\alpha_{sm}\, \Lambda^2}}\right)\right]~,
\eeq
is expected to be much larger,\cite{Eichten:1983hw} due to the factor
of $g^2/4\pi\alpha_{sm}$.

\end{enumerate}

By searching for deviations in four-fermion processes from
the predictions of the standard model, we can place
bounds on possible contact interactions. In principle, any interactions
of the form 
\beq
{\cal L}_{FF}={g^2\over 2!\,\Lambda^2}\, \eta_{ LL} 
(\bar{\psi}_L \gamma^\mu \psi_L)(\bar{\psi}_L \gamma_\mu \psi_L) 
+ ({ RR,\,LR})
\eeq
consistent with chiral symmetry, gauge symmetries, and flavor symmetries may
be present. The convention followed in the typical analyses is {$\alpha =
  g^2/4\pi =1$}. Note that this is lower than would be expected from
dimensional analysis and tends to {\it understate the limits on $\Lambda$}.
In addition, the analyses generally set only one coefficient ({\it e.g.}
$\eta_{LL}$ or $\eta_{LR}$) to be non-zero at a time, and give it a value of
$\pm 1$, which effectively ``normalizes'' $\Lambda$.  Current lower bounds on
the scale $\Lambda$ are shown in table \ref{compositeness}, and are typically
several TeV.\cite{Groom:2000in}

\begin{table}
\begin{center}
\begin{tabular}{|c|c|}
\hline
$\Lambda^+_{LL}(eeee)>\, 3.1\, {\rm TeV}$ & {\tt OPAL}\\
$\Lambda^-_{LL}(eeee)>\, 3.8\, {\rm TeV}$& \\
\hline
$\Lambda^+_{LL}(ee\mu\mu)>\, 4.5\, {\rm TeV}$ & {\tt OPAL}\\
$\Lambda^-_{LL}(ee\mu\mu)>\, 4.3\, {\rm TeV}$& \\
\hline
$\Lambda^+_{LL}(ee\tau\tau)>\, 3.8\, {\rm TeV}$& {\tt OPAL}\\
$\Lambda^-_{LL}(eeee)>\, 4.0\, {\rm TeV}$& \\
\hline
$\Lambda^+_{LL}(\ell\ell\ell\ell)>\, 5.2\, {\rm TeV}$& {\tt OPAL}\\
$\Lambda^-_{LL}(\ell\ell\ell\ell)>\, 5.3\, {\rm TeV}$& \\
\hline
$\Lambda^+_{LL}(eeqq)>\, 4.4\, {\rm TeV}$& {\tt OPAL}\\
$\Lambda^-_{LL}(eeqq)>\, 2.8\, {\rm TeV}$&\\
\hline
$\Lambda^+_{LL}(\nu\nu qq)>\, 5.0\, {\rm TeV}$& {\tt NUTEV}\\
$\Lambda^-_{LL}(\nu\nu qq)>\, 5.4\, {\rm TeV}$& \\
\hline
$\Lambda^\pm_{LL}(qqqq)>\, 1.9\, {\rm TeV}$&{\tt D{\O}}\\
\hline
\end{tabular}
\end{center}
\caption{Current limits on the scale $\Lambda$ for compositeness derived from
various four-fermion scattering experiments.\protect\cite{Groom:2000in}}
\label{compositeness}
\end{table}

\subsection{Gauge Bosons}

Next we consider the possibility that the ordinary gauge bosons are
composite objects. At first sight this seems unreasonable, since these
particles are associated with a local gauge symmetry. However, as
shown by Weinberg,\cite{Weinberg:1964ew} for consistency {\it any} massless
vector particle must couple to a {\it conserved} current -- {\it i.e.}
the existence of a ``gauge symmetry'' is automatic for any massless
spin-1 particle.

Dimensional analysis allows us to estimate the size of the resulting
couplings. The natural size of the coupling constant is { ${\cal O}(4 \pi)$},
as can be seen by considering a generic three-point coupling:
\beq
{{ \Lambda^4}\over{ g^2}}\, 
\left({\partial \over { \Lambda}}\right)
\left({{ g}V \over { \Lambda}}\right)^3
\equiv
g (\partial V) V V~.
\eeq
However, the standard model gauge interactions are asymptotically free
and their couplings at high energies ($\simeq 1$ TeV) are small. For
this reason, it is likely that the $\gamma$, $g$, $W^\pm_T$, $Z_T$ are
fundamental.

The situation is quite different for the longitudinally polarized weak
gauge bosons, the $W^\pm_L$ and $Z_L$. These particles are ``eaten''
Goldstone bosons, have effective couplings to each other proportional
to momentum and, as discussed in the previous lecture, they are not
fundamental in theories of dynamical electroweak symmetry breaking.
The effective lagrangian for a theory of massive electroweak bosons
with composite longitudinal modes includes the fundamental $W_T$ and
$B_T$ gauge bosons of $SU(2)_W \times U(1)_Y$ symmetry. In addition,
just as in the chiral lagrangian in QCD, we may describe the Goldstone
bosons of electroweak symmetry breaking by a matrix $\Sigma$ which
transforms to $L \Sigma R^\dagger$ under a global $SU(2)_L \times
SU(2)_R$ which is broken to $SU(2)_V$.  As discussed previously, the
residual ``custodial'' $SU(2)_V$ symmetry
ensures\cite{Weinstein:1973gj,Sikivie:1980hm} that the weak
interaction $\rho$ parameter (eq. \ref{custodial}) is equal to 1.

The low-energy of effects of dynamical electroweak symmetry breaking
include anomalous weak gauge-boson couplings described by the
effective lagrangian described above. These corrections can be thought
of as due to the exchange of the lightest resonances likely to be
present in such theories, ``technirho'' vector mesons ($\rho_{TC}$)
analogous to the $\rho$ in QCD. Such corrections modify the 3-pt
functions\footnote{There are also ``vacuum polarization'' corrections
  to the 2-pt functions, generally expressed in terms of contributions
  to the
  Peskin-Takeuchi\protect\cite{Peskin:1990zt,Peskin:1992sw,Golden:1991ig,Holdom:1990tc,Dobado:1991zh}
  $S$ and $T$ parameters.\protect\cite{Lane:1993wz}}:
\beq
{\lower15pt\hbox{\epsfysize=0.75 truein \epsfbox{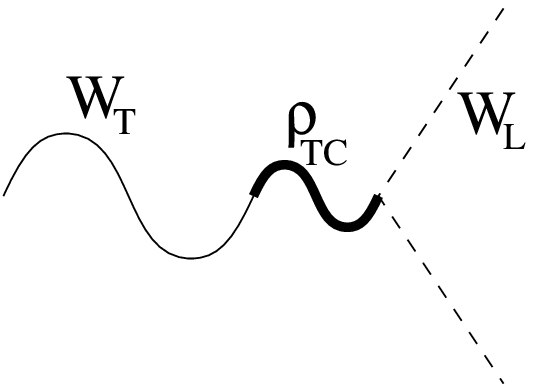}}}
\eeq
yielding the couplings
\beq
 -\ i\, { g_{SU(2)}} {{\it l}_{9L} \over 16 \pi^2}\, \tr {{ \W^{\mu \nu}} D_\mu
\Sigma D_\nu \Sigma^\dagger}~,
\eeq
and
\beq
-\ i\, { g'_{U(1)}} {{\it l}_{9R} \over 16 \pi^2}\, \tr {{ B^{\mu \nu}}
D_\mu \Sigma^\dagger D_\nu\Sigma}~.
\eeq
In these expressions we have taken ${ g} \simeq 4\pi$, so the {\it l}'s
are normalized to be {\cal O}(1).

\begin{figure}[thb]
\begin{center}
\includegraphics*[width=8cm]{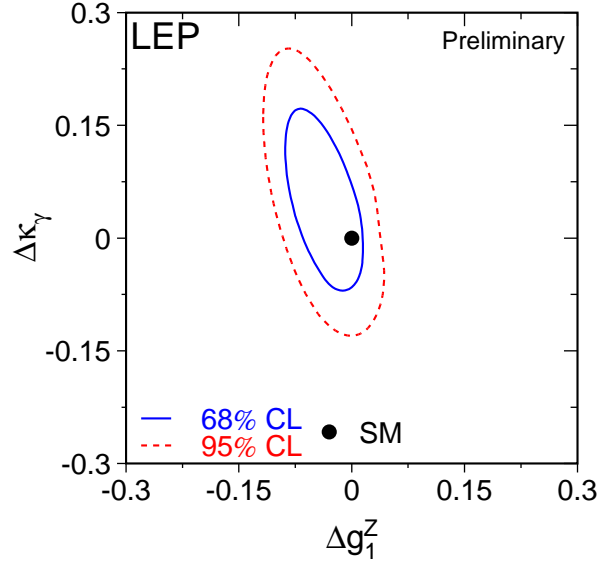}
\end{center}
\label{leptgc}
\caption{Current limits\protect\cite{LEPEWWGTGC:1999} on the anomalous weak gauge-boson coupling
parameters $\kappa_\gamma$ and $g^Z_1$.}
\end{figure}

The conventional\cite{Hagiwara:1987vm} description of anomalous weak
gauge-boson couplings was given by Hagiwara, {\it et. al.}:
\beqa
{i\over e \cot\theta} & {\cal L}_{WWZ}  =  g_1
(W^\dagger_{\mu\nu} W^\mu Z^\nu - W^\dagger_\mu Z_\nu W^{\mu\nu}) \\
& + \kappa_Z W^\dagger_\mu W_\nu Z^{\mu\nu} + {\lambda_Z\over
M_W^2}W^\dagger_{\lambda\mu}W^\mu_\nu Z^{\nu\lambda}~, \nonumber
\eeqa
and
\beqa
{i\over e}  & {\cal L_{WW\gamma}} =  (W^\dagger_{\mu\nu} W^\mu A^\nu -
W_\mu^\dagger A_\nu W^{\mu\nu}) \\
&  + \kappa_\gamma W^\dagger_\mu W_\nu
F^{\mu\nu} + {\lambda_\gamma\over M_W^2} W^\dagger_{\lambda\mu}W^\mu_\nu
F^{\nu\lambda}~.\nonumber
\eeqa
Comparing with the interactions above (in unitary gauge, $\Sigma \equiv {\cal I}$),
we find:
\beq
\left.
\begin{array}{c}
g_1 - 1\\
\kappa_Z - 1\\
\kappa_\gamma - 1
\end{array}
\right\}
\ \approx\ {\alpha_* l_i \over 4\pi \sin^2\theta} = {\cal O}(10^{-2}-10^{-3})~.
\eeq
The couplings $\lambda_{Z,\gamma}$ arise from higher order
interactions, and are estimated to be ${\cal O}(10^{-4}-10^{-5})$.
Current limits from LEP are shown in fig. 2.  These data
show agreement with the standard model, {but do not set useful
  limits}.

\bigskip
\noindent\underline{\tt For your consideration\ldots}
\bigskip

\begin{enumerate}

  
\item If quarks and leptons are composite, one expects there are excited
  states with the same quantum numbers (typically denoted $\ell^*$ and
  $q^*$). The PDG\cite{Groom:2000in} lists bounds on excited states of quarks
  and leptons of ${\cal O}(100\, {\rm GeV})$.

\begin{itemize}
\item Based on dimensional analysis,\footnote{See also, Weinberg and
    Witten.\protect\cite{Weinberg:1980kq}} what bound does this place
  on the scale of compositeness $\Lambda$?

\end{itemize}

\item BNL experiment E-821 will measure the anomalous magnetic moment
  of the muon to 0.35 {\it ppm}. 

\begin{itemize}

\item Show using dimensional analysis that
  this is measurement should be sensitive to one-loop weak
  corrections.  

\item If the experiment agrees with the SM, what bound will
  this measurement place on the scale of muon compositeness?

\end{itemize}

\end{enumerate}

\section{Composite Higgs Bosons}

Up to now, our discussion of compositeness has consisted of the
construction of consistent effective {\it low-energy} theories and an
analysis of current experimental lower bounds on the scale of
compositeness of the observed particles. In order to proceed, we need
to understand the characteristics of plausible fundamental {\it
  high-energy} theories which give rise to light (possibly massless)
composite states. In this section we consider models which give rise
to composite scalars, and in particular a Higgs Boson.  In subsequent
sections, we will describe models giving rise to composite fermions
and gauge bosons.

\subsection{{Top-Condensate Models}}

The fact that the top quark, with $m_t \simeq 2M_W,\, 2M_Z$, is much
heavier than other fermions implies that the top is more strongly
coupled to the EWSB sector. This has led to the construction\footnote{The
  phenomenology of these models, especially as they relate to the top
  quark, is discussed in the lectures by Elizabeth Simmons in this
  volume.} of models in which
all\cite{Miranskii:1989ds,Miranskii:1989xi,Nambu:1989jt,Marciano:1989xd,Bardeen:1990ds,Hill:1991at,Cvetic:1997eb,Dobrescu:1998nm,Chivukula:1998wd}
or some\cite{Hill:1995hp} of electroweak symmetry breaking is due to
top condensation, $\langle \bar{t}t\rangle \neq 0$.

\begin{figure}
\begin{center}
\begin{minipage}[t]{5.5cm}
\includegraphics*[width=5.5cm]{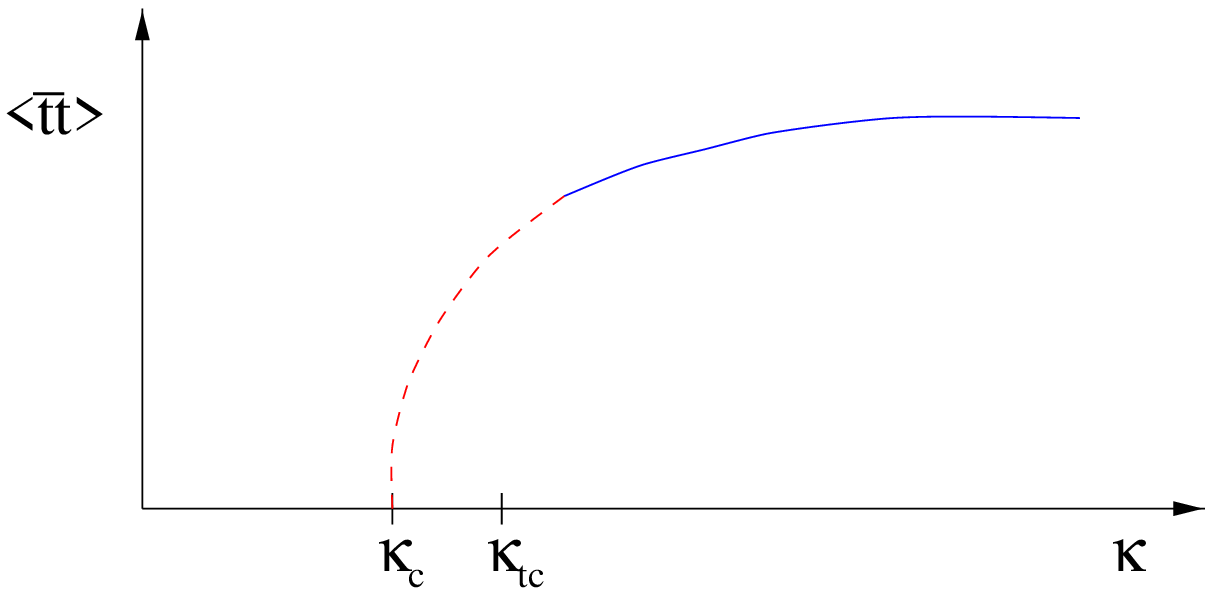}
\caption {Top-quark condensate produced in NJL approximation to
topcolor dynamics as a function of the coupling $\kappa$. Condensate
forms only for $\kappa > \kappa_c$ and, at least in this approximation,
the condensate turns on continuously -- {\it i.e.} the quantum chiral
phase transition is second order.}
\label{critical}
\end{minipage}
\hspace{2mm}
\begin{minipage}[t]{5.5cm}
\includegraphics*[width=5.5cm]{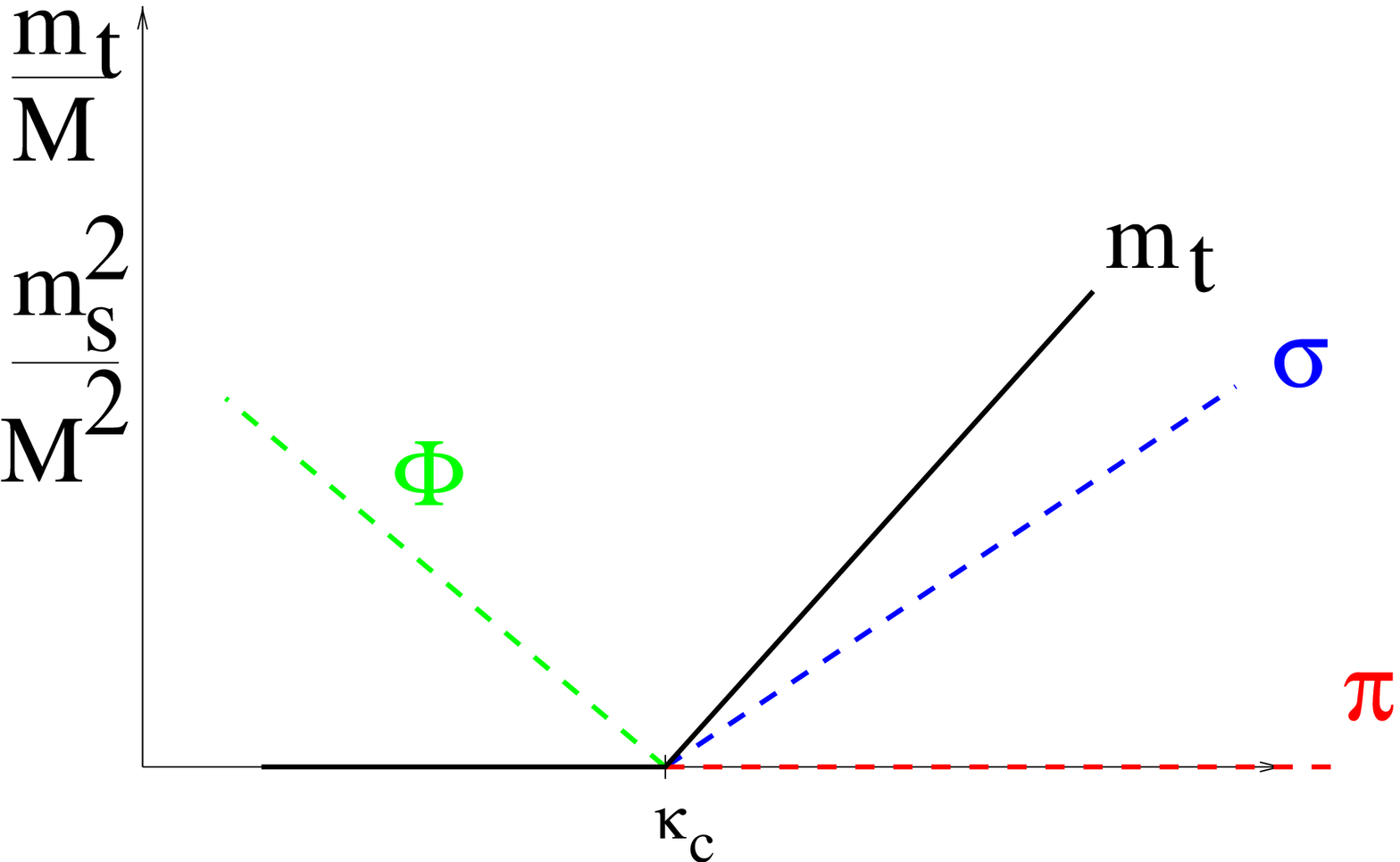}
\caption{Spectrum of low-energy states in {\it any} model with a 
{\it second order} quantum chiral phase transition. The spectrum must
change smoothly at the critical coupling; there must be massless fermions
below the critical coupling and massless Goldstone bosons above the
critical coupling.}
\label{second}
\end{minipage}
\end{center}
\end{figure}

The simplest model involves a spontaneously broken but
strong topcolor\cite{Hill:1991at} gauge interaction
\beq
{ SU(3)_{tc}} \times { SU(3)}
\stackrel{M}{\to} SU(3)_{QCD}
\eeq
which couples preferentially to the third generation of quarks.
At energies small compared to the mass ($M$) of the topgluon,
such an interaction gives rise to a local four-fermion
operator
\beq
{\cal L} \supset - {4\pi\kappa \over 2!\, M^2}
\left(\overline{Q}\gamma_\mu {\lambda^a\over 2} Q\right)^2
\eeq
where $\kappa \propto g^2_{tc}(M)$, the $\lambda^a$ are the Gell-mann
matrices, and the fields $Q$ are the doublets (left- and right-handed,
for the model as described so far) third generation weak doublet
fields. 

This model may be solved\cite{Nambu:1961er} in the ``NJL
approximation'' in the large-$N_c$ limit. The behavior of the chiral
symmetry breaking condensate is shown in fig. \ref{critical}, with:
\beq
{\langle\bar{t}t\rangle \over M^3}
\propto \sqrt{\Delta\kappa \over \kappa_c} \equiv
\sqrt{{\kappa-\kappa_c}\over \kappa_c}
~.
\eeq
The condensate changes smoothly from zero as $\kappa$ exceeds the critical
value $\kappa_c$; this behavior represents a second order chiral phase
transition.  Clearly, if $M \gg 1\tev$ and the dynamics of topcolor occurs at
scales much higher than 1 TeV, the value of $\kappa$ must be tuned close to
$\kappa_c$.

Assuming the transition is second order, as motivated by the NJL
calculation, it is easy to understand the form of the effective
low-energy field theory when $\kappa$ is just slightly greater than
$\kappa_c$. As shown in fig. \ref{second}, the light degrees of
freedom include the top-quark, the Goldstone bosons eaten by the $W$
and $Z$, as well as a singlet scalar particle (the $\sigma$). In order
for this theory to have a smooth limit as $\kappa \to \kappa^+_c$, the
$\sigma$ and the Goldstone bosons must arrange themselves to form a
light composite Higgs boson! 

In fact, for the theory as described, we have not distinguished the
top from the bottom quark. The theory includes two light Higgs
doublets in a $2\times 2$ matrix field $\Phi$ as required in an
$SU(2)_L \times SU(2)_R$ linear sigma model. Using our rules of
dimensional analysis, the most general effective lagrangian describing
the light fields is:
\beq
{\cal L}_{eff} = {\rm Tr}\left(\partial^\mu\Phi^\dagger \partial_\mu\Phi\right) 
+ \bar{\psi}\slashchar{\partial} \psi
+ y\, \bar{\psi}\Phi \psi + {\cal O}\left({\partial^2 \over M^2}\right)~.
\eeq
Note that, the absence of a Higgs mass term ($m^2 {\rm Tr}(\Phi^\dagger
\Phi)$) is due entirely to the {\it dynamical assumption} that we are (very)
close to the transition ($|m^2| \ll M^2$). Dimensional analysis implies
that $y = {\cal O}(4\pi)$ and a heavy top quark arises naturally.

The Higgs (at large $N_c$ in the NJL approximation) can be found
directly as pole in the sum of bubble sum diagrams\cite{Bardeen:1990ds}
\beq
\epsfxsize=5cm\epsffile{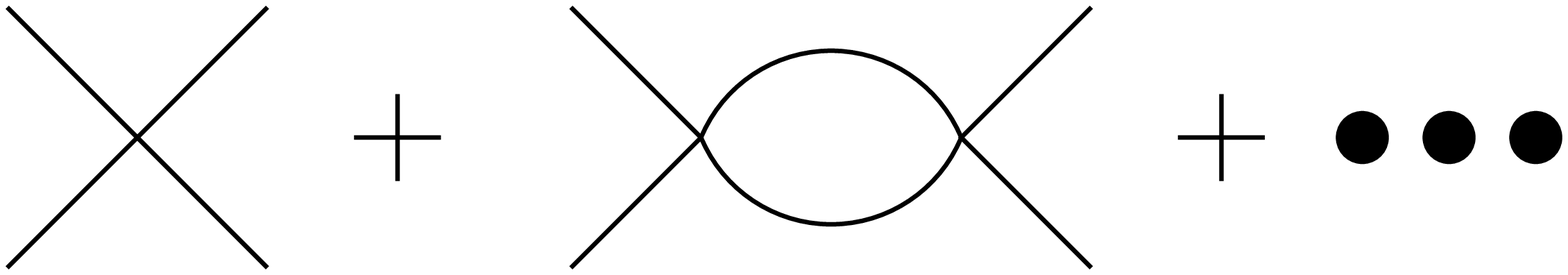}~.
\eeq
The eaten Goldstone bosons arise in the corresponding diagrams for the
$W/Z$ self-energies:
\beq
\epsfxsize=7cm\epsffile{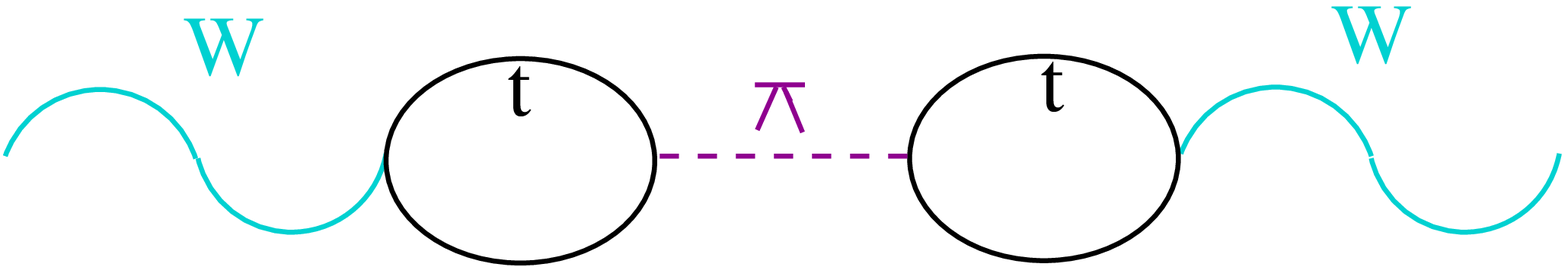}~,
\eeq
and implies a Higgs vacuum expectation value
\beq
f_t^2 = {N_c \over 4 \pi} \int^\Lambda {k^2 dk^2 m_t^4 
\over (k^2 + m_t^2)^2}~.
\eeq

A number of phenomenological issues must be addressed prior to
constructing a realistic model based on topcolor.  First, additional
``tilting'' interactions must be introduced to ensure that
the bottom quark is not heavy, {\it i.e.} $\langle \bar{b} b \rangle
\approx 0$. Second, some account must be given of the observed mixing
between the third generation and the first two. Finally, top quark
condensation alone produces only a Higgs vacuum expectation value of
$f_t \approx 60$ GeV, which is too small to account for electroweak
symmetry breaking.\cite{Bardeen:1990ds}

The simplest model of a single composite Higgs boson based
on top-condensation is the top seesaw model.\cite{Dobrescu:1998nm,Chivukula:1998wd}
In this model, electroweak symmetry breaking is due to the condensate
of the left-handed top quark $t_L$ with a new right-handed weak singlet quark
$\chi_R$. While $\langle t_L \chi_R\rangle$ is responsible for
all of electroweak symmetry breaking, mixing of the top with left- and
right-handed singlet quarks yields a seesaw mass matrix
\beq
{\lower55pt\hbox{\includegraphics*[width=5cm]{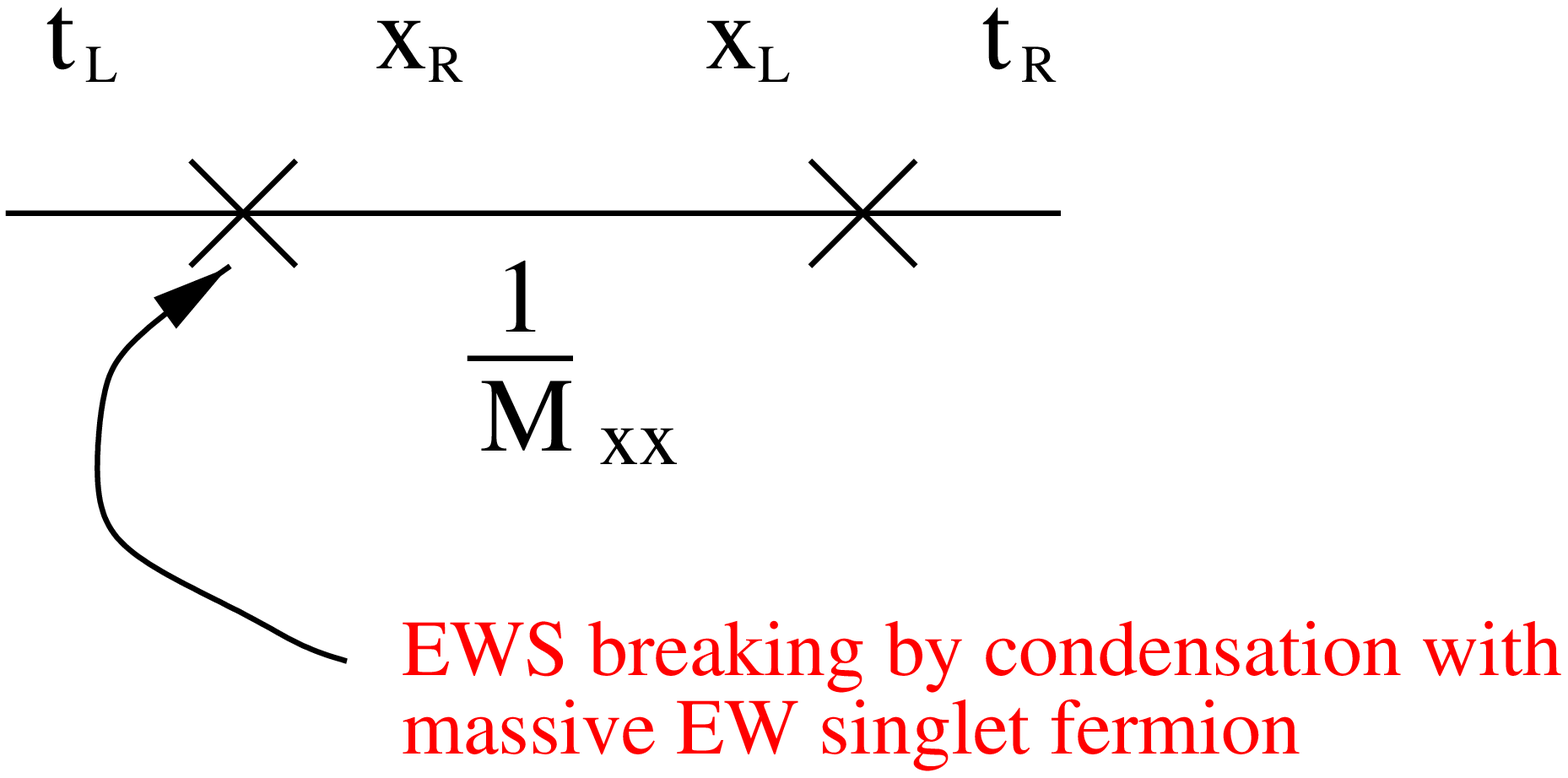}}}
\hskip+10pt{
\pmatrix{\overline{t_L}&\overline{\chi_L}\cr }\,
\pmatrix{0& m_{t\chi}\cr \mu_{\chi t}&\mu_{\chi\chi}\cr}
\pmatrix{t_R\cr \chi_R\cr}}
\eeq
which gives rise to the observed mass-eigenstate top-quark.

\subsection{The Triviality of the Standard Higgs Model\protect\footnote{The
work presented in this and the following two subsections has appeared
previously.\protect\cite{Chivukula:2000fe}}}

A composite Higgs is also motivated by the fact that the standard
one-doublet Higgs model {\it does not strictly exist} as a continuum
field theory. This result is most easily illustrated in terms of the
Wilson renormalization
group.\cite{Wilson:1971bg,Wilson:1971dh,Wilson:1974jj} Any quantum
field theory is defined using a regularization procedure which
ameliorates the bad short-distance behavior of the theory.  Following
Wilson, we define the scalar sector of the standard model
\beqa
{\cal L}_\Lambda =  & D^\mu \phi^\dagger D_\mu \phi + 
m^2(\Lambda)\phi^\dagger \phi 
+ {\lambda(\Lambda)\over 4}(\phi^\dagger\phi)^2 \\
& + {\eta(\Lambda)\over 36\Lambda^2}(\phi^\dagger\phi)^3+\ldots  
\nonumber
\eeqa
in terms of a {fixed} UV-cutoff $\Lambda$. Here we have allowed for
the possibility of terms of (engineering) dimension greater than four.
While there are an infinite number of such terms, one representative
term of this sort, $(\phi^\dagger \phi)^3$, has been included
explicitly for the purposes of illustration. Note that the coefficient
of the higher dimension terms includes the appropriate number of
powers of $\Lambda$, the intrinsic scale at which the theory is
defined.

\begin{figure}
\begin{center}
\begin{minipage}[t]{5.5cm}
\includegraphics*[bb=100 95 640 515,width=5.5cm]{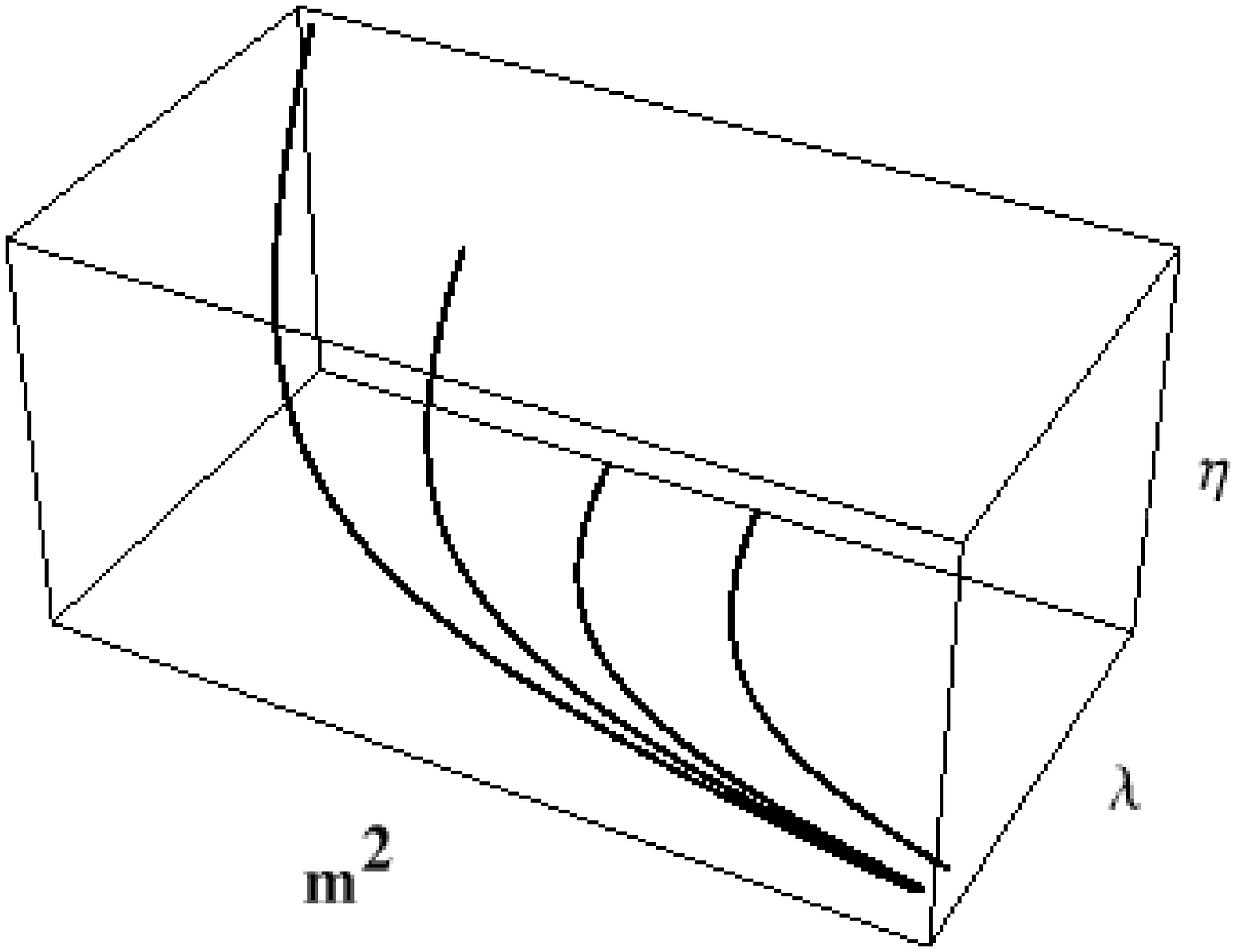}
\caption {Graphical representation of Wilson RG flow of 
  $(m^2(\Lambda),\lambda(\Lambda),\eta(\Lambda))$. As we scale to low
  energies, $m^2 \to \infty$, $\lambda \to 0$, and $\eta \to 0$.}
\label{3drg}
\end{minipage}
\hspace{2mm}
\begin{minipage}[t]{5.5cm}
\includegraphics*[width=5.5cm]{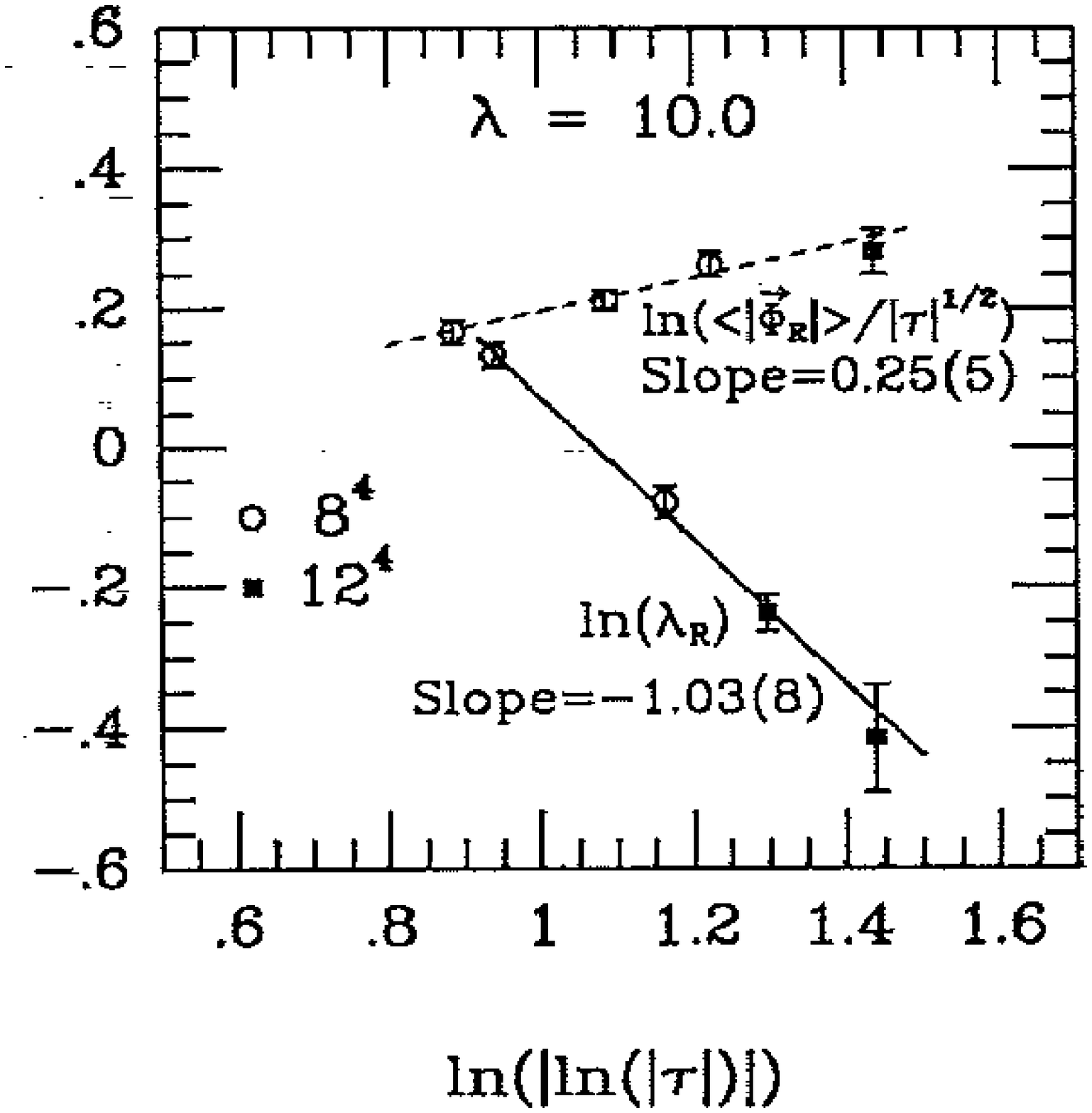}
\caption{Results of a nonperturbative lattice monte carlo 
  study\,\protect\cite{Kuti:1988nr} of the scalar sector of the standard model
  with bare coupling $\lambda=10$. The approximate slope of -1 for the
  renormalized coupling, $\lambda_R$, shows agreement with the naive
  one-loop perturbative result.}
\label{kutishenfig}
\end{minipage}
\end{center}
\end{figure}

Wilson observed that, for the purposes of describing experiments
at some fixed low-energy scale $E \ll \Lambda$, it is possible to trade
a high-energy cutoff $\Lambda$ for one that is slightly
lower, $\Lambda^\prime$, so long as $E \ll \Lambda^\prime < \Lambda$.
In order to keep low-energy measurements fixed, it will in
general be necessary to redefine the values of the coupling constants
that appear in the Lagrangian. Formally, this process
is referred to as ``integrating out'' the
(off-shell) intermediate states with $\Lambda^\prime < k < \Lambda$. Keeping
the low-energy properties fixed we find
\beqa
{\cal L}_\Lambda & \Rightarrow & {\cal L}_{\Lambda^\prime} \nonumber\\
m^2(\Lambda)& \rightarrow & m^2(\Lambda^\prime) \\
\lambda(\Lambda) & \rightarrow & \lambda(\Lambda^\prime) \nonumber \\
\eta(\Lambda) & \rightarrow & \eta(\Lambda^\prime)  ~.
\nonumber
\eeqa

Wilson's insight was to see that many properties of the theory can be
summarized in terms of the evolution of these (generalized) couplings
as we move to lower energies. Truncating the infinite-dimensional
coupling constant space to the three couplings shown above, the
behavior of the scalar sector of the standard model is illustrated in
Figure \ref{3drg}. This figure illustrates a number of important
features of scalar field theory. As we flow to the infrared, {\it
  i.e.}  lower the effective cutoff, we find:
\begin{itemize}
\item $\eta \to 0$ --- this is the modern interpretation of
renormalizability. If $m_H \ll \Lambda$, the theory is drawn to
the two-dimensional $(m_H, \lambda)$ subspace. Any theory
in which $m_H \ll \Lambda$ is therefore close to a renormalizable theory with
corrections suppressed by powers of $\Lambda$.
\item $m^2 \to \infty$ --- This is the naturalness or hierarchy problem.
To maintain $m_H \simeq {\cal O}(v)$ we must adjust\,\footnote{Nothing we
discuss here will address the hierarchy problem directly.} the value
of $m_H$ in the underlying theory to of order
\beq
{\Delta m^2(\Lambda) \over m^2(\Lambda)} \propto {v^2 \over \Lambda^2}~.
\eeq
\item $\lambda \to 0$ --- The coupling $\lambda$ has a positive
  $\beta$ function and, therefore, as we scale to low energies
  $\lambda$ tends to 0.  If we try to take the ``continuum'' limit,
  $\Lambda \to +\infty$, the theory becomes free or
  trivial.\cite{Wilson:1971bg,Wilson:1971dh,Wilson:1974jj}
\end{itemize}

The triviality of the scalar sector of the standard one-doublet Higgs
model implies that this theory is only an effective low-energy theory
valid below some cut-off scale $\Lambda$.  Given a value of {$m^2_H =
  2 \lambda(m_H) v^2$}, there is an {\it upper} bound on {$\Lambda$}.
An {\it estimate} of this bound can be obtained by integrating the {one-loop} $\beta$-function, which yields
\beq
\Lambda \laem m_H \exp\left({4\pi^2v^2\over 3m^2_H}\right)~.
\label{landau}
\eeq
For a light Higgs, the bound above is at uninterestingly high scales
and the effects of the underlying dynamics can be too small to be
phenomenologically relevant. For a Higgs mass of order a few hundred
GeV, however, effects from the underlying physics can become
important. I will refer to these theories generically as ``composite
Higgs'' models.

Finally, while the estimate above is based on a perturbative analysis,
nonperturbative investigations of $\lambda \phi^4$ theory on the
lattice show the same behavior. This is illustrated in Figure
\ref{kutishenfig}.

\subsection{$T$, $S$, and $U$ in Composite Higgs Models}

In an $SU(2)_W \times U(1)_Y$ invariant scalar theory of a single doublet,
all interactions of dimension less than or equal to four also respect a
larger ``custodial'' symmetry\,\cite{Weinstein:1973gj,Sikivie:1980hm} which
insures the tree-level relation $\rho=M^2_W / M^2_Z \cos^2\theta_W\equiv 1$
is satisfied. The leading custodial-symmetry violating operator is of
dimension six\,\cite{Buchmuller:1986jz,Grinstein:1991cd} and involves four
Higgs doublet fields $\phi$. In general, the underlying theory does not
respect the larger custodial symmetry, and we expect the interaction
\beq
{\lower35pt\hbox{\epsfysize=1.00 truein \epsfbox{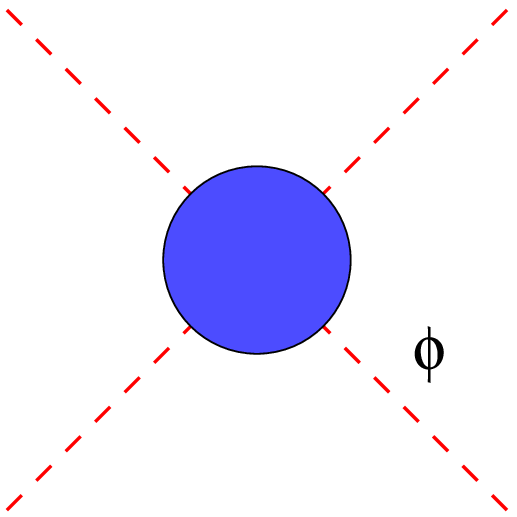}}}
\Rightarrow
{b { \kappa}^2 \over 2!\, { \Lambda}^2} 
(\phi^\dagger \stackrel{\leftrightarrow}{D^\mu} \phi)^2~,
\label{toperator}
\eeq
to appear in the low-energy effective theory.  Here $b$ is an unknown
coefficient of ${\cal O}(1)$, and ${\kappa}$ measures size of
couplings of the composite Higgs field. In a strongly-interacting
theory, $\kappa$ is expected\,\cite{Manohar:1984md,Georgi:1993dw} to be of ${\cal O}(4\pi)$.

\begin{figure}
\begin{center}
\begin{minipage}[t]{5.5cm}
\includegraphics[width=5.5cm]{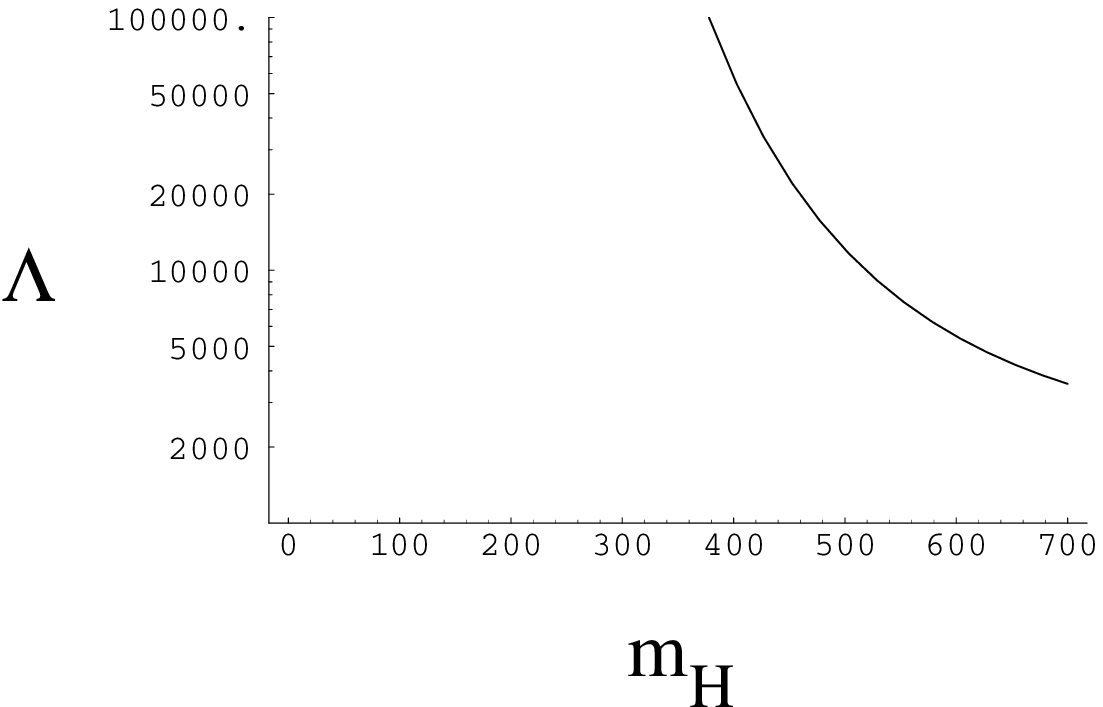}
\caption{Upper bound on scale $\Lambda$ as per eq. (\protect\ref{landau}).}
\label{landaugraph}
\end{minipage}
\hspace{2mm}
\begin{minipage}[t]{5.5cm}
\includegraphics[width=5.5cm]{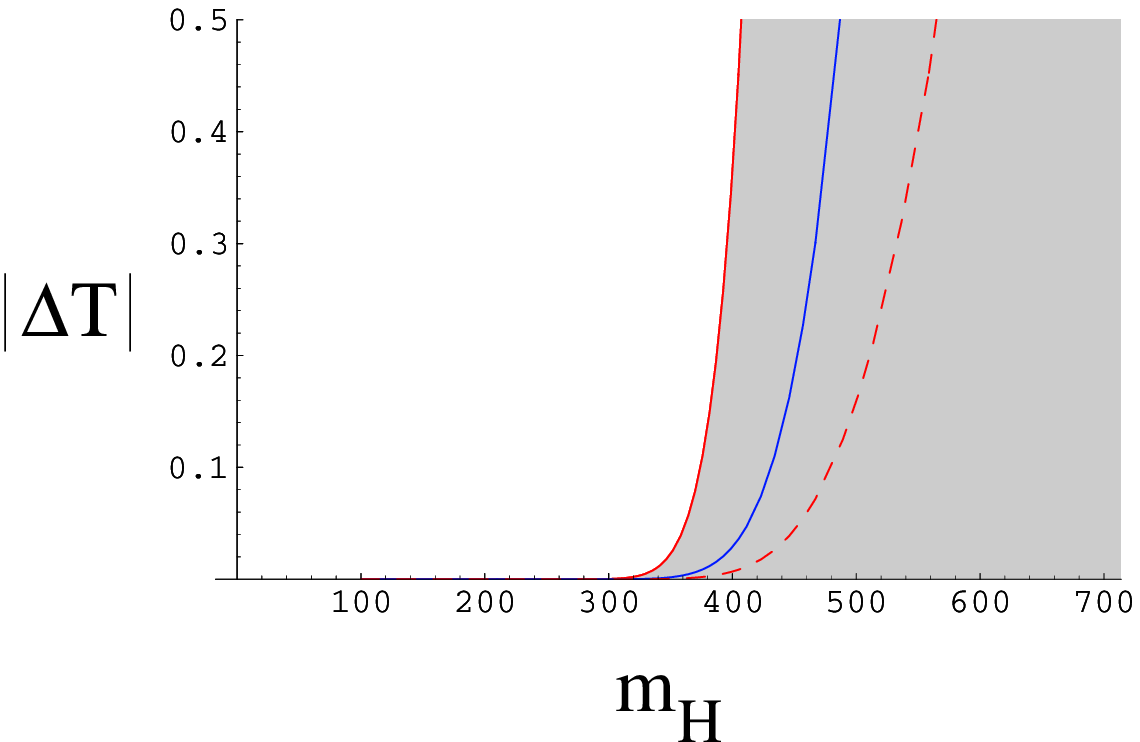}
\caption{Lower bound on expected size of $|\Delta T|$ as per eq. (\protect\ref{tbound}),
for $|b|\kappa^2=16\pi^2$, $4\pi$, and 3.}
\label{tboundgraph}
\end{minipage}
\end{center}
\end{figure}

Deviations in the low-energy theory from the standard model can be
summarized in terms of the ``oblique''
parameters\,\cite{Peskin:1990zt,Peskin:1992sw,Golden:1991ig,Holdom:1990tc,Dobado:1991zh}
$S$, $T$, and $U$.  The operator in eq. \ref{toperator} will give
rise to a deviation ($\Delta \rho= \varepsilon_1 = \alpha T$)
\beq { |\Delta T|} = { |b| {\kappa}^2 {v^2 \over \alpha(M_Z)
    { \Lambda}^2}} { \gaem} {|b|\kappa^2\, v^2 \over \alpha(M^2_Z)\, m^2_H}\,
\exp\left({-\,{8 \pi^2 v^2\over 3 m^2_H}}\right) ~, 
\label{tbound}
\eeq
where $v \approx 246$ GeV and we have used eq. \ref{landau} to obtain the
final inequality.  The consequences of eqns. (\ref{landau}) and
(\ref{tbound}) are summarized in Figures \ref{landaugraph} and
\ref{tboundgraph}. The larger $m_H$, the lower $\Lambda$ and the larger the
expected value of $\Delta T$.  Current limits imply $|T| \stackrel{<}{\sim}
0.5$, and hence\cite{Chivukula:1996sn} ${ \Lambda \stackrel{>}{\sim} 4\, {\rm
    TeV} \cdot \kappa}$.  (For $\kappa \simeq 4\pi$, {$m_H \laem 450$ GeV}.)

By contrast, the leading contribution to $S$ arises from
\beq
{\lower35pt\hbox{\epsfysize=1.00 truein \epsfbox{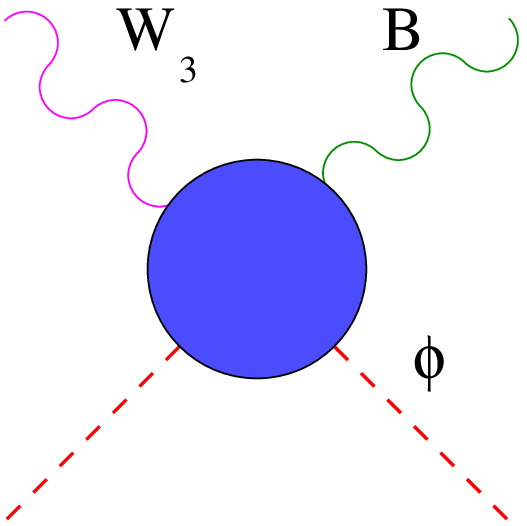}}}
\Rightarrow
-\,{a \over 2!\,{ \Lambda}^2} \left\{ [D_\mu,D_\nu] \phi \right\}^\dagger[D^\mu,D^\nu]\phi~.
\eeq
This gives rise to ($\varepsilon_3={\alpha S/4\sin^2\theta_W}$)
\beq \Delta S = {4 \pi a  v^2\over  { \Lambda}^2}~. \eeq
It is important to note that the size of contributions to $\Delta T$
and $\Delta S$ are very different
\beq 
{\Delta S \over \Delta T} = {a \over b} \left({4\pi \alpha \over { \kappa}^2}\right) ={\cal
  O}\left({10^{-1}\over { \kappa}^2}\right)~.
\eeq
Even for ${\kappa}\simeq 1$, $|\Delta S| \ll |\Delta T|$.

Finally, contributions to $U$ ($\varepsilon_2=-{\alpha U\over 4\sin^2\theta_W}$),
arise from
\beq
{c g^2 {\kappa^2}\over \Lambda^4} (\phi^\dagger W^{\mu\nu}\phi)^2
\eeq
and, being suppressed by $\Lambda^4$,  are typically much smaller than $\Delta T$.

\subsection{Limits on a Composite Higgs Boson}

\begin{figure}
\begin{center}
\includegraphics[width=8cm]{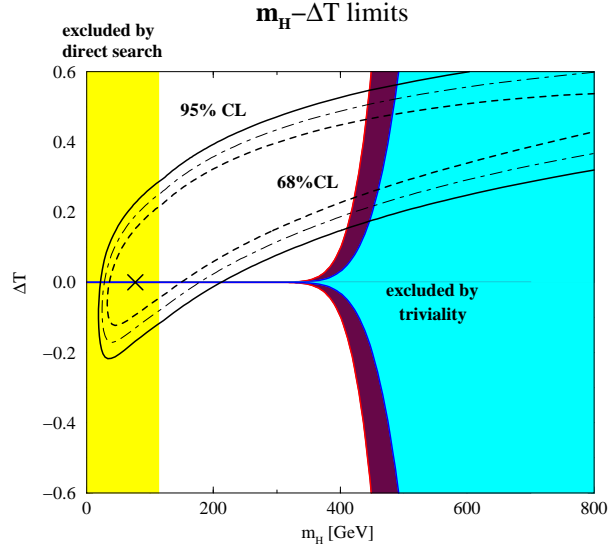}
\caption{68\% and 95\% CL regions allowed\,\protect\cite{Chivukula:2000px} in $(m_H,|\Delta T|)$
  plane by precision electroweak
  data.\protect\cite{LEPEWWG:2000} Fit allows for $m_t$,
  $\alpha_s$, and $\alpha_{em}$ to vary consistent with current
  limits.\protect\cite{Chivukula:2000px} Also shown by the dot-dash curve is the
  contour corresponding to $\Delta \chi^2=4$, whose intersection with
  the line $\Delta T=0$ -- at approximately 190 GeV -- corresponds to
  the usual 95\% CL upper bound quoted on the Higgs boson mass in the
  standard model. The triviality bound curves are for
  $|b|\kappa^2=4\pi$ and $4\pi^2$, corresponding to representative
  models.\protect\cite{Chivukula:2000px}}
\label{mht_zfitter}
\end{center}
\end{figure}

From triviality, we see that the Higgs model can only be an effective
theory valid below some high-energy scale $\Lambda$.  As the Higgs
becomes heavier, the scale $\Lambda$ {\it decreases}. Hence, the
expected size of contributions to $T$ {\it grow}, and are larger than
the expected contribution to $S$ or $U$. The limits from precision
electroweak data in the $(m_H, \Delta T)$ plane are shown in Figure
\ref{mht_zfitter}.  We see that, for positive $\Delta T$ at 95\% CL,
the allowed values of the Higgs mass extend to well beyond 800 GeV. On the
other hand, not all values can be realized consistent with the bound
given in eq.  (\ref{landau}). As shown in figure \ref{mht_zfitter},
values of the Higgs mass beyond approximately 500 GeV would likely require
values of $\Delta T$ much larger than allowed by current measurements.

I should emphasize that these estimates are based on dimensional arguments. I
am not arguing that it is {\it impossible} to construct a composite Higgs
model consistent with precision electroweak tests with $m_H$ greater than 500
GeV.  Rather, barring accidental cancellations in a theory without a
custodial symmetry, contributions to $\Delta T$ consistent with eq.
\ref{landau} are generally to be expected. In particular composite Higgs
boson models, the bounds given here have been shown to
apply.\cite{Chivukula:2000px}

\begin{figure}
\begin{center}
\begin{minipage}[t]{5.5cm}
\includegraphics[width=5.5cm]{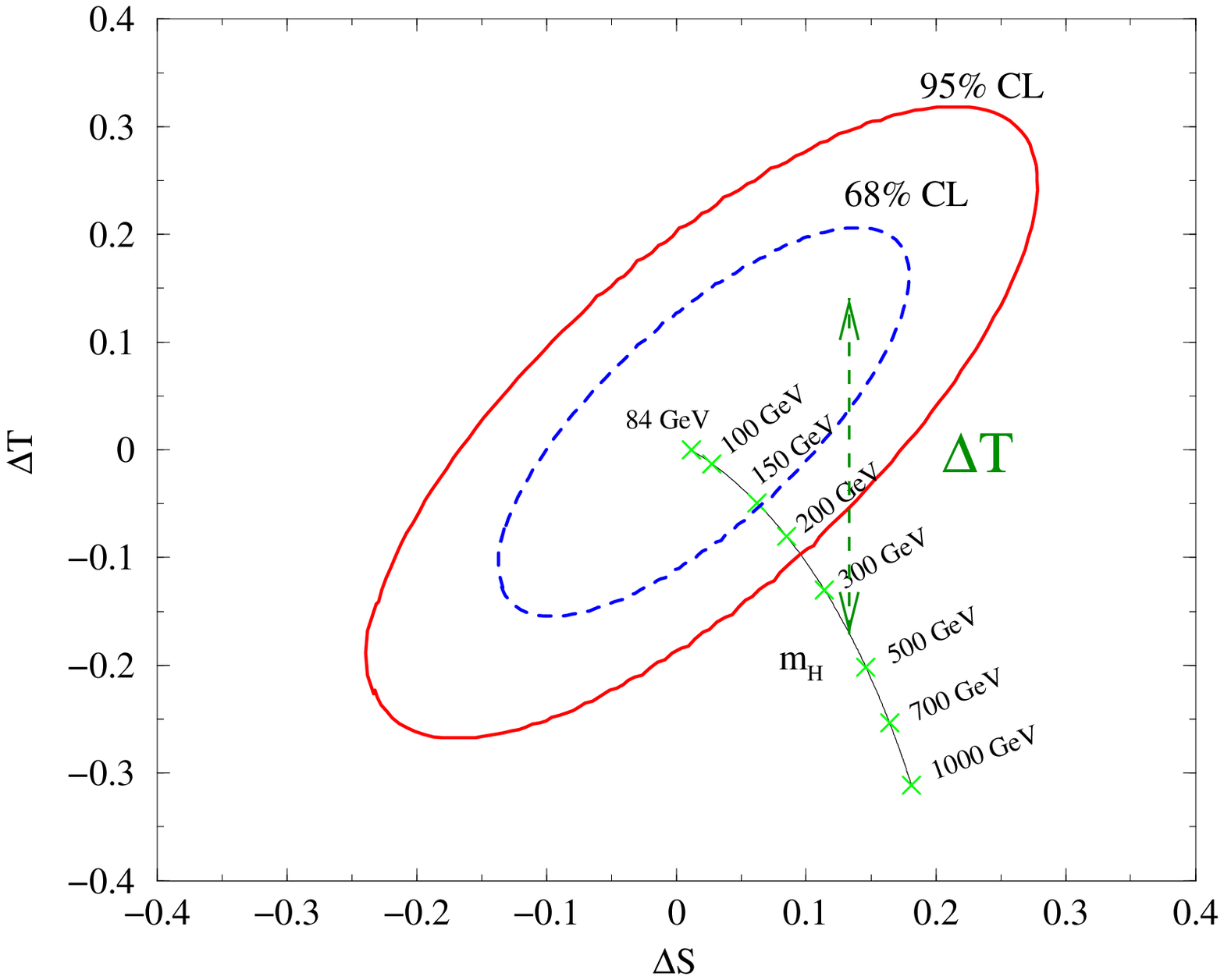}
\caption{68\% and 95\% CL regions allowed in $(\Delta S, \Delta T)$
plane by precision electroweak data.\protect\cite{LEPEWWG:2000}
Fit allows for $m_t$, $\alpha_s$, and $\alpha_{em}$ to vary consistent
with current limits.\protect\cite{Chivukula:2000px} Standard model prediction for
varying Higgs boson mass shown as parametric curve, with $m_H$ varying from
84 to 1000 GeV.}
\label{stplot_varymt}
\end{minipage}
\hspace{2mm}
\begin{minipage}[t]{5.5cm}
\includegraphics[width=5.5cm]{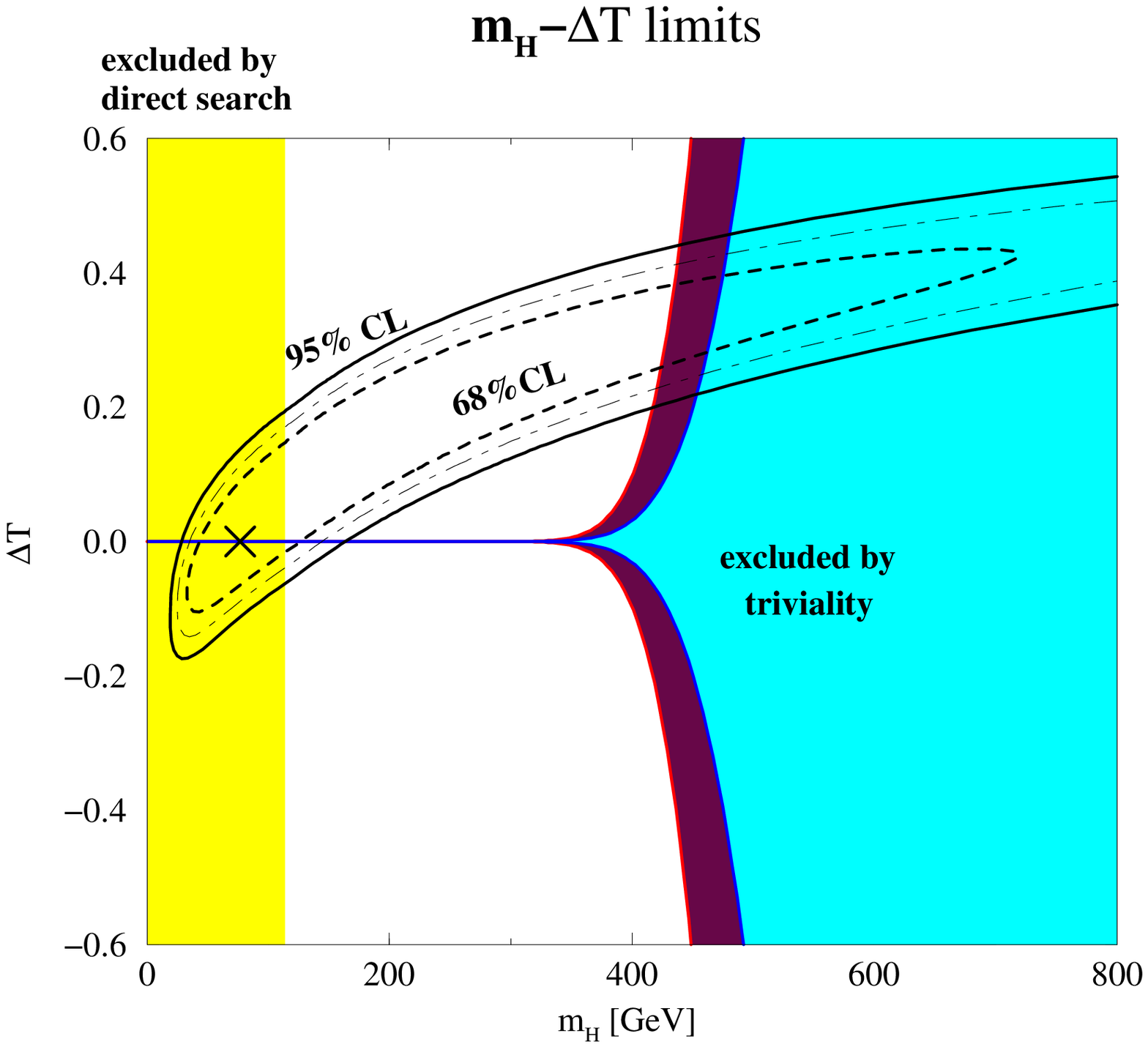}
\caption{68\% and 95\% CL allowed region\,\protect\cite{Chivukula:2000px} 
  in $(m_H,\Delta T)$ plane if Fermilab Tevatron Run II reduces the
  uncertainty in top-quark and W-boson masses to $\Delta m_t = 2$ GeV and
  $\Delta M_W = 30$ MeV about their current central values.}
\label{mht_future}
\end{minipage}
\end{center}
\end{figure}

These results may also be understood by considering limits in the
$(S,T)$ plane for {\it fixed} $(m_H,m_t)$. In Figure
\ref{stplot_varymt}, changes from the nominal standard model best fit
($m_H=84$ GeV) value of the Higgs mass are displayed as contributions
to $\Delta S(m_H)$ and $\Delta T(m_H)$. Also shown are the 68\% and
95\% CL bounds on $\Delta S$ and $\Delta T$ consistent with current
data. We see that, for $m_H$ greater than ${\cal O}$(200 GeV), a
positive contribution to $T$ can bring the model within the allowed
region.

At Run II of the Fermilab Tevatron, it may be possible to reduce the
uncertainties in the top-quark and W-boson masses to $\Delta m_t = 2$
GeV and $\Delta M_W = 30$ MeV.\cite{Runii:1996}  Assuming that the measured values of
$m_t$ and $M_W$ equal their current central values, such a reduction
in uncertainties will result the limits in the $(m_H, \Delta T)$ plane
shown in Figure \ref{mht_future}. Note that, despite reduced
uncertainties, a Higgs mass of up to 500 GeV or so will still
be allowed.

\section{Composite Fermions}

\subsection{Chiral Symmetry and Anomalies}

Consider a chiral transformation of a 4-component Dirac field:
\beq
\psi \to e^{+i\alpha \gamma_5} \psi \hskip+10pt \Leftrightarrow
\hskip+10pt \psi_{L,R} \to e^{\pm i \alpha} \psi_{L,R}~,
\eeq
which may also be written 
\beq
\psi_L \to e^{+i\alpha} \psi_L \hskip+10pt \& \hskip+10pt
{\cal C}\psi_R \equiv \psi^c_L \to e^{+i\alpha}\psi^c_L~,
\eeq
in terms of the charge-conjugate field $\psi^c_L$. Since fermion
mass terms 
\beq
m\bar{\psi}\psi \hskip+10pt \Leftrightarrow \hskip+10pt
m\bar{\psi_L}\psi_R + {h.c.}
\eeq
couple the left- and right-handed components of a fermion field, mass
terms are not chirally invariant. Therefore, an unbroken global chiral
symmetry is a {\it sufficient} condition for the existence of massless
fermions at low energies. If these massless fermions are composite
objects, the fundamental fermions of which they are composed must
carry the same symmetries. 't Hooft realized\cite{'tHooft:1980xb} that
there were additional constraints relating the {\it representations}
of the fundamental and composite fermions: the anomaly matching
conditions.

We begin our discussion with a review of anomalies in quantum field
theory. Regularization of quantum field theory, necessary to extract
finite answers, generally breaks various global symmetries of the
model, and one must re-impose these symmetries upon renormalization.
However, regularization always breaks chiral symmetries and,
surprisingly, we generally {\it cannot} re-impose
both\cite{Bell:1969ts,Adler:1969gk,Jackiw:1969xt,Schwinger:1951nm}
vector and chiral symmetries in renormalized theory! The result is
that a symmetry of the classical theory is broken at quantum level.
Hence, the use of the term anomaly.

In perturbation theory, the anomaly manifests itself in the
behavior of the triangle diagram
\beq
\lower20pt\hbox{\includegraphics*[width=3cm]{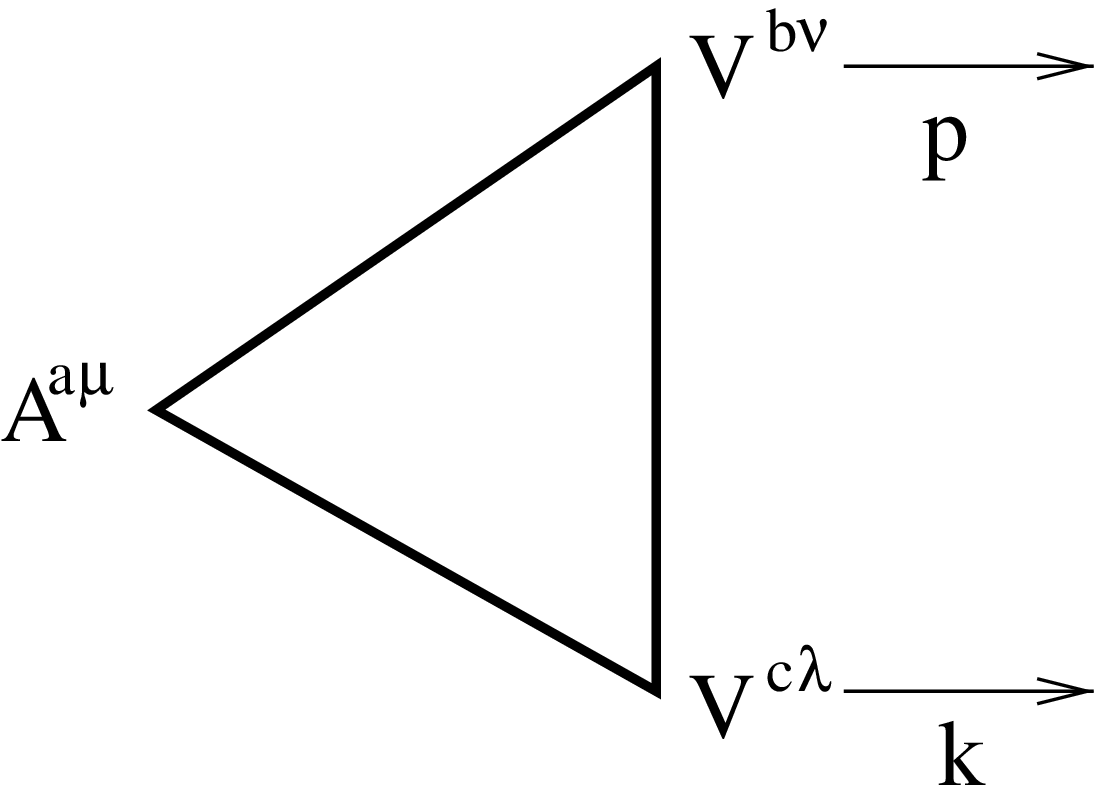}}~.
\eeq
Imposing vector current conservation, one finds that the divergence of the
axial current is nonzero:
\beq
\langle p,\nu,b;k,\lambda,c|\partial_\mu A^{a\mu} | 0\rangle
 =  {1\over 4\pi^2}\, \varepsilon^{\alpha\nu\beta\lambda}\,p_\alpha k_\beta\,
\cdot 
{\rm Tr}(T^a\{ T^b,T^c \})~.
\eeq
For simplicity, we will write the theory, and symmetries, in terms of
left-handed fermions and currents only. Diagrammatically, one can move the
chiral projector $\left({{1-\gamma_5}\over 2}\right)$ to a single
vertex. We will regularize so that the resulting VVV term not
anomalous; the AVV term remains as above.

\bigskip
\noindent\underline{\tt (Gauge)$^3$ Anomalies}
\bigskip

Consider first a vectorial $SU(N_c)$ gauge theory with
$N_f$ fermions in the fundamental $N_c$ representation. 
The fermions transform as:
\beq
\psi_L:\, (N_c, N_f, 1) \hskip+15pt \psi^c_L:\, (\bar{N_c},1,\bar{N_f})
\eeq
under the $SU(N_c)$ gauge and $SU(N_f)_L \times SU(N_f)_R$ global symmetries.
For consistency, the $SU(N_c)$ gauge current must be {\it conserved} and the
corresponding symmetry cannot be anomalous. For a representation $R$, define:
\beq
{\rm Tr}(T^a_R\{T^b_R,T^c_R\}) \equiv {1\over 2}\, {\cal A}(R)\, d^{abc}~,
\label{anomaly}
\eeq
and hence ${\cal A}(N_c) \equiv 1$. For the $\bar{N_c}$ representation:
\beq
{\rm Tr}(T^a_R\{T^b_R,T^c_R\}) = {\rm Tr}((-T^a)^T\{(-T^b)^T,(-T^c)^T\})~,
\eeq
and ${\cal A}(\bar{N_c}) = -1$. Hence the total gauge anomaly
\beq
N_f\cdot {\cal A}(N_c) + N_f \cdot {\cal A}(\bar{N_c}) \equiv 0~.
\eeq
In general, for {\it any} real representation the generators $T^a_R$
are unitarily equivalent to $(-T^a)^T$ and therefore ${\cal
  A}(R)\equiv 0$.

One can also construct a chiral gauge theory using a complex, but
anomaly-free, representation. Consider the two-index antisymmetric tensor
representation, $A^{ij}$, of $SU(N)$.  Since $d_{abc}$ is a group-theoretic
invariant, to calculate ${\cal A}(R)$ one need only consider a single
nonvanishing term on the left hand side of eq. \ref{anomaly}. In particular,
consider the generator proportional to
\beq 
\left(
\begin{array}{cccc}
-1 & & & \\
 &-1 & & \\
 &   & \ddots & \\
&&& N-1
\end{array}
\right)
\label{generator}
\eeq
in the fundamental representation. ${\rm Tr}(T^3)$ implies
\beq
{\cal A}(N) \propto (N-1)\cdot(-1)^3 + 1\cdot(N-1)^3
\eeq
for the fundamental representation $N$ and
\beq
{\cal A}(A) \propto {(N-1)(N-2)\over 2}\cdot (-2)^3 + (N-1)\cdot (N-2)^3
\eeq
for the antisymmetric tensor representation $A$. A little algebra shows that
${\cal A}(A) = (N-4)\, {\cal A}(N)$, hence we can construct an anomaly-free
chiral gauge theory by including fermions transforming as one antisymmetric
tensor and $N-4$ antifundamentals.

\bigskip
\noindent\underline{\tt For your consideration \ldots}
\bigskip

Consider one family of quarks and leptons in
the standard $SU(3)_C \times SU(2)_W \times U(1)_Y$
model.

Show that all of the following gauge anomalies cancel:

\begin{itemize}

\item $(SU(3)_C)^3$

\item $(SU(2)_W)^3$

\item $(U(1)_Y)^3$

\item $(SU(3)_C)^2\, SU(2)_W$

\item $(SU(2)_W)^2\, SU(3)_C$

\item $(SU(3)_C)^2\, U(1)_Y$

\item $(SU(2)_W)^2\, U(1)_Y$

\item $(U(1)_Y)^2\, SU(3)_C$

\item $(U(1)_Y)^2\, SU(2)_W$

\item $SU(3)_C\, SU(2)_W\, U(1)_Y$

\item { $\{SU(3)_C,SU(2)_W,U(1)_Y\}\, (Grav)^2$}

\end{itemize}

{\it Which connect quark and lepton charges?}

\bigskip
\noindent\underline{\tt Global/Gauge Anomalies}
\bigskip

Global chiral symmetries can be violated by anomalies as well.
The classic example is axial quark number, $U(1)_A$, in QCD.
Including this (approximate) classical global symmetry the
$N_f$ quarks transform as
\beq
\psi^i_L:\, (3,N_f,1)_{+1} \hskip+20pt \psi^c_L:\, (\bar{3},1,\bar{N_f})_{+1}
\eeq
under $SU(3) \times SU(N_f)_L \times SU(N_f)_R \times U(1)_A$.
The triangle graph
\beq
{\lower20pt\hbox{\includegraphics*[width=2cm]{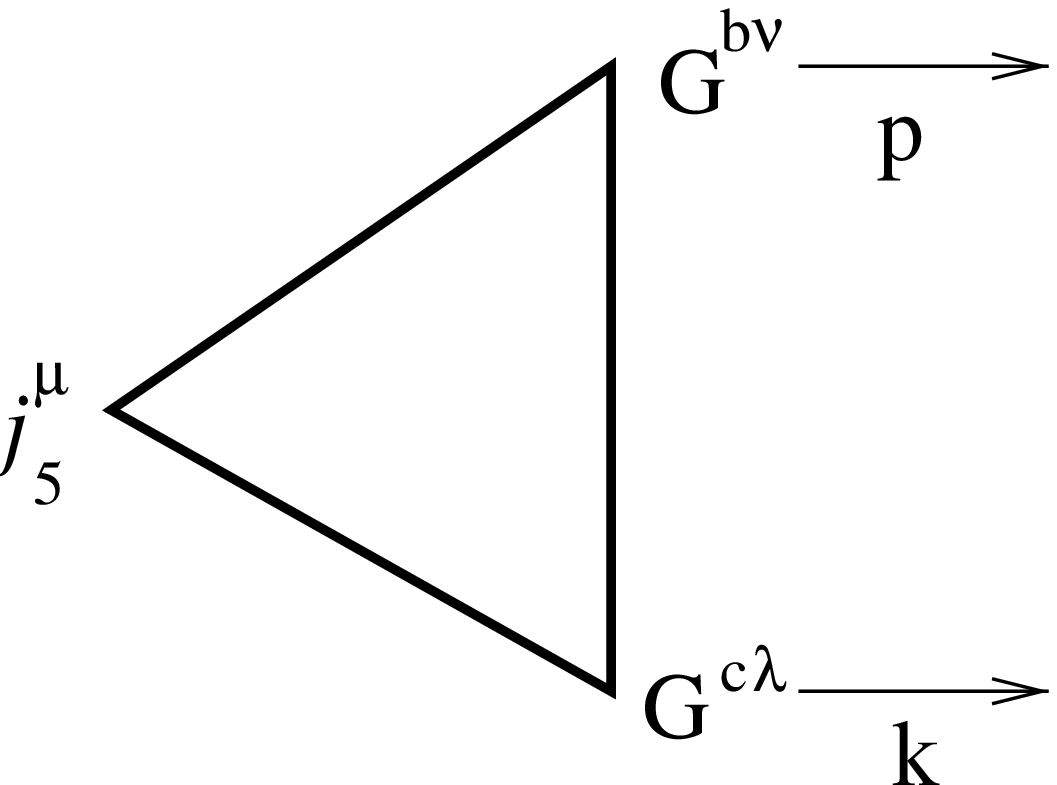}}}
\eeq
yields a result proportional to
\beq
{\rm Tr}(T_{U(1)_A} \{T^b,T^c\})= N_f\, {\rm Tr}\{T_3^b,T_3^c\} + 
N_f\, {\rm Tr}\{T_{\bar{3}}^b,T_{\bar{3}}^c\}\neq 0~.
\eeq
Consequently,\cite{'tHooft:1976up} the $U(1)_A$ current is not
conserved
\beq
{ \partial^\mu j^5_\mu = -\, {g_{QCD}^2 N_f\over 32\pi^2} 
\varepsilon^{\alpha\nu\beta\lambda} F^a_{\alpha\nu} F^a_{\beta\lambda}}~,
\eeq
and there is no ninth Goldstone Boson in QCD!

Surprisingly, there is an anomaly-free global $U(1)$ symmetry in the
chiral $SU(N)$ model described above. The $U(1)_{global} SU(N)^2$
anomaly is proportional to index
\beq
{\rm Tr}\{T^a_R,T^b_R\} \equiv k(R)\, \delta^{ab}~.
\eeq
of the $SU(N)$ representation of the fermion, which is 
a group theoretic invariant. Consider generator proportional to
\beq
\left(
\begin{array}{ccc}
1 & & \\
 &-1 & \\
 &   & \ddots \\
\end{array}
\right)~.
\eeq
The index of a representation is proportional to ${\rm Tr}(T^2)$,
which gives
\beq
k(N) \propto 2\cdot (1)^2 + (N-2)\cdot (0)^2
\label{indexfund}
\eeq
for the fundamental representation, and
\beq
k(A) \propto 1\cdot(0)^2 + 2 \cdot (N-2)\cdot(1)^2
\label{indexanti}
\eeq
for the antisymmetric tensor. Recall that the consistent chiral $SU(N)$ theory has
one antisymmetric tensor $A$ and $N-4$ antifundamental representations
$\bar{N}$.  Comparing to eqs. \ref{indexfund} and \ref{indexanti}, we
see that there is an anomaly-free global $U(1)$ symmetry under
which\footnote{Our choice of the $U(1)$ charges allows this symmetry
  to commute with the global $SU(N-4)$ symmetry on the $N-4$
  antifundamental fields.} the antisymmetric tensor has charge $-1$
and $(N-4)$ antifundamentals have charge ${{N-2\over N-4}}$.  In
the simplest nontrivial case, chiral $SU(5)$, the antisymmetric tensor
is ten dimensional and the fermion fields transform as $10_{-1}$ and
$\bar{5}_3$ under $SU(5) \times U(1)$.

\subsection{{(Global)$^3$ Anomalies:} the `t Hooft Conditions}

The existence of massless composite fermions implies that there is a
low-energy global chiral symmetry group $H$, and this group must be a {\it
  subgroup} of the high-energy global symmetries.  `t Hooft
argued\cite{'tHooft:1980xb} that the (global)$^3$ anomaly factor (${\cal
  A}_H$) {\it must be the same} in the low- and high-energy theories. His
argument runs as follows: consider a theory with massless composite fermions
with a global chiral symmetry group $H$.  Suppose you were to {\it weakly}
gauge the chiral global symmetry group $H$. In order to avoid gauge
anomalies, one must also add ``spectator'' fermions which are weakly, but not
strongly interacting, to cancel the anomalies ${\cal A}_H$. By definition,
these weak gauge interactions don't affect the dynamics, and the massless
composite fermions must still form. In order for the weak gauge group to
remain consistent, therefore the low-energy massless composites must cancel
anomalies of spectator fermions. Hence,
\beq
{\cal A}_{fundamental} = {\cal  A}_{composite}~.
\label{matching}
\eeq
The condition in eq. \ref{matching} is a nontrivial relation between
the $H$ representations of the fundamental and composite fermions.  It
provides a {\it necessary} condition which must be satisfied by any
putative theory of composite fermions. We will illustrate the anomaly
matching conditions with two plausible theories of composite massless
fermions.

Consider the chiral $SU(5)$ gauge theory described earlier with fundamental
fermion transforming as $10_{-1}$ ($\chi^{ij}$) and $\bar{5}_3$ ($\psi_{k}$).
As discussed, the theory has a global chiral $U(1)$ free of strong anomalies
with the charges shown.  It is possible to construct an $SU(5)$ singlet
fermion: $\chi^{{ i}{ j}} \psi_{i} \psi_{ j}$. The $U(1)$ charge of this
composite fermion is: -1+3+3=5. Therefore 
\beq
{\cal A}_{composite} =(5)^3 = 125 = {\cal A}_{fundamental}= 10\cdot(-1)^3 +
5\cdot(3)^3~. 
\eeq
Assuming $SU(5)$ confines, it is possible that the chiral $U(1)$ symmetry is
unbroken and the a single massless fermion is present in the low-energy
spectrum.

In the case of $SU(5)$ there is a complementary picture of the physics
which, surprisingly, yields the same low-energy spectrum.
Consider the possible bilinear (scalar) condensates of fermions:
\beq
\epsilon^{\alpha\beta} \psi_{\alpha i}\psi_{\beta j} = (\overline{15})_6~,
\eeq
\beq
\epsilon^{\alpha\beta} \psi_{\alpha i}\chi^{jk}_\beta = (5 + 45)_2~,
\eeq
and
\beq
\epsilon^{\alpha\beta} \chi^{il}_\alpha\chi^{jk}_\beta = (\bar{5}+50)_{-2}~.
\eeq
Of these four channels we can guess\footnote{for a review of the most
  attractive channel hypothesis, and chiral symmetry breaking in
  general, see Michael Peskin's 1982 lectures at the Les Houches
  summer school.\protect\cite{Peskin:1982mu}} that the most
attractive 
\beq
\includegraphics*[width=3cm]{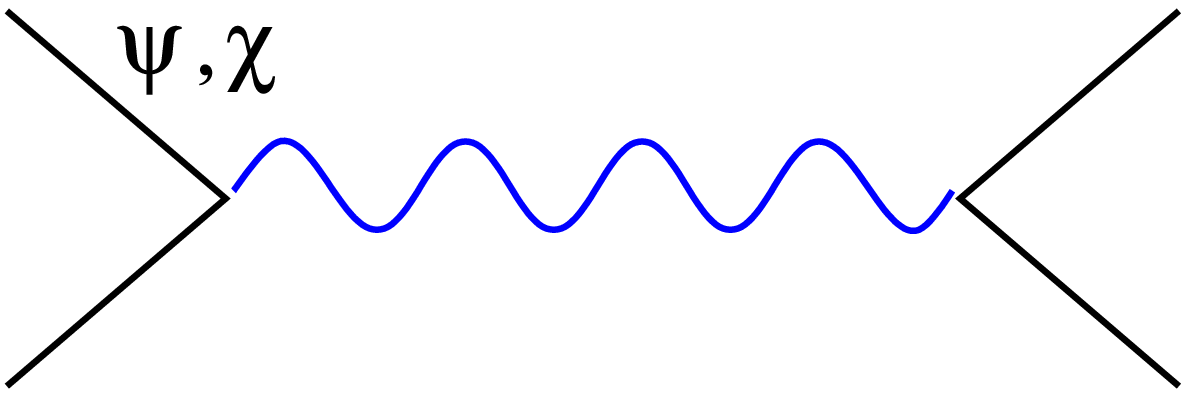}
\eeq
two-fermion channel forms first. The diagram above is proportional
to
\beq
T^a_A \cdot T^a_B = {1\over 2}\left((T^a_A + T^a_B)^2 - (T^a_A)^2 
- (T^a_B)^2\right) \propto {\cal C}(A+B) - {\cal C}(A)-{\cal C}(B)
\label{mac}
\eeq
where ${\cal C}(A)$ is the Casimir of representation $A$. The most
attractive channels (corresponding to the most negative value of the
expression in eq. \ref{mac}) are the smallest representations: $5_2$
+ $\bar{5}_{-2}$.

Assuming that these two $5_{\pm 2}$ condensates form, and that their
vevs align, we find the symmetry breaking pattern $SU(5)_{gauge}
\times U(1)_{global} \to SU(4)_{gauge} \times \tilde{U}(1)_{global}$.
The residual global $\tilde{U}(1)$ is a combination of the original global
$U(1)$ charge with the diagonal $SU(5)$ generator, $\tilde{Q}=(2Q+Q_5)/5$
with
\beq
Q_5= \left(
\begin{array}{cc|c}
1 & &\\
& \ddots & \\
\hline 
& & -4 \\
\end{array}
\right)~.
\eeq
Under the unbroken symmetry, the original fermion representations decompose
as $\bar{5}_{3} \to \bar{4}_1 + 1_2$ and $10_{-1} \to 6_0 + 4_{-1}$. The
residual $SU(4)$ gauge symmetry is vectorial, and we expect condensates to
give dynamical masses corresponding to the condensates $\bar{4}\cdot 4$ and
$6 \cdot 6$. Both of these condensates carry zero global charge, and
$\tilde{U}(1)$ remains unbroken. Therefore, the gauge singlet fermion
($1_{2}$) remains massless!

{\it Both the confining picture and the gauge-symmetry-breaking /
  Higgs phase picture have the same low-energy spectrum.} These two
pictures are {\it complementary}. Indeed, the correspondence can be
seen directly in terms of the fields: $\chi \psi \psi \Leftrightarrow
\langle \chi \psi\rangle \psi$, where we explicitly note the {\it
  condensate} in which the singlet fermion propagates.  Note that the
global symmetry ($U(1)$) is the same in the two pictures, while the gauge
symmetry is not. This is consistent because a gauge symmetry is a
redundancy in the Hilbert space, while a global symmetry relates {\it
  different} physical states.

\begin{figure}
\begin{center}
\includegraphics*[width=5cm]{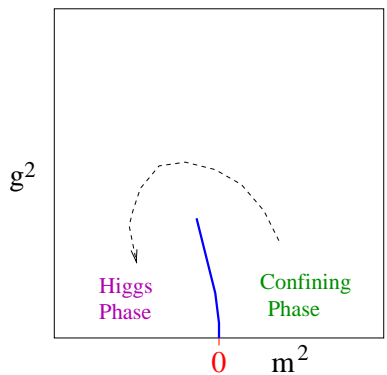}
\end{center}
\caption{Complementarity: In a gauge theory (coupling $g$) with scalars 
  (mass $m^2$) in the fundamental representation, Fradkin and Shenker
  have shown\protect\cite{Fradkin:1979dv} that the confining and Higgs
  phases can be smoothly connected. This implies that the {\it
    massless} spectrum is the same in both phases.}
\label{complementarity}
\end{figure}

Note that, in the Higgs phase picture, the condensate $\langle \chi
\psi\rangle$ transforms in the {\it fundamental} representation
of the gauge group. Fradkin and Shenker have
shown\cite{Fradkin:1979dv} that in a gauge theory with scalars
transforming in the fundamental representation, the confining and
Higgs phases are smoothly connected (see fig. \ref{complementarity}).
This implies that the {\it massless} spectrum is the same in both
phases. The behavior we have noted in the chiral $SU(5)$ gauge theory
is a dynamical realization of complementarity.

\subsection{Mooses\protect\cite{Georgi:1986hf,Georgi:1986dw}}
  
Georgi has proposed a class of composite models with QCD-like
dynamics. The models are most easily described diagrammatically
(``moose'' diagrams), with the basic element
\beq
\includegraphics*[width=2cm]{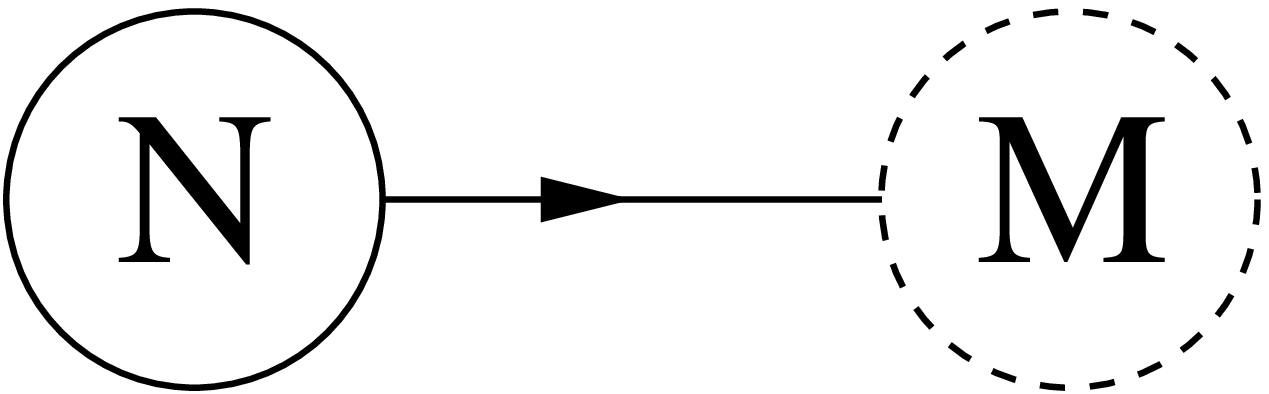}
\eeq
where the solid circle denotes an $SU(N)$ gauge group, the dashed
circle an $SU(M)$ global group, and the line a left-handed
$(N,\bar{M})$ fermion. In this notation, QCD with three light flavors
is
\beq
\includegraphics*[width=3.5cm]{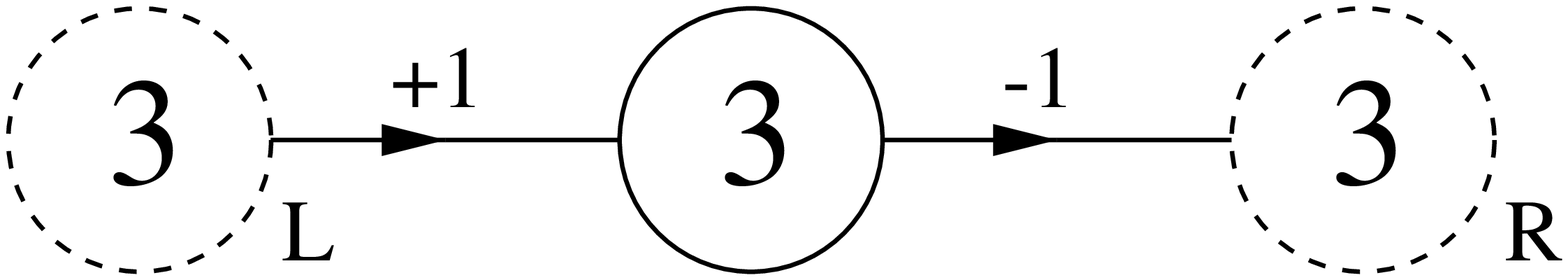}
\eeq
where the global $SU(3)_L\times SU(3)_R$ are shown by the outer
circles and $SU(3)_C$ by the middle circle. The fermions
$(3,\bar{3},1)_{+1}$ are denoted by the left-hand line, and
$(1,3,\bar{3})_{-1}$ by the right-hand one. Finally, the charges for
the non-anomalous $U(1)_B$ are shown by the numbers above the line.
The constraints of anomaly cancellation are easily seen: gauge
anomalies are canceled whenever the number of lines leaving a solid
circle equal the number of lines entering; nonanomalous global
$U(1)$'s exist if the total charge of the fermions coupled to a given
gauge group equals zero. After QCD chiral symmetry breaking, the
fermions condense and the residual global symmetries ``collapse'' the
moose diagram to
\beq
\includegraphics*[width=0.60cm]{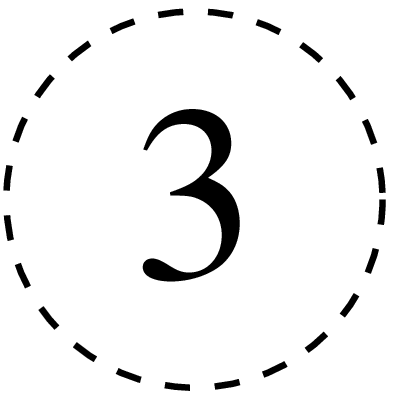}
\eeq
which denotes the residual vector $SU(3)$ symmetry.

The simplest nontrivial model of this sort is the
``odd linear moose''
\beq
\includegraphics*[width=5cm]{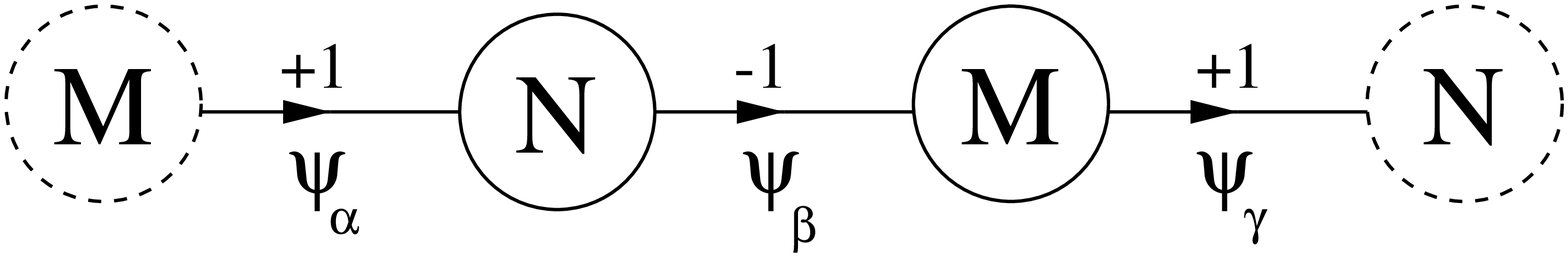}
\eeq
which has an $SU(M) \times SU(N) \times U(1)$ global symmetry.
Assuming both gauge groups confine, all global anomalies are saturated
by a massless $\psi_\alpha \psi_\beta\psi_\gamma$ bound state with
quantum numbers $(M,\bar{N})_{+1}$.

The odd linear moose has {\it two} complementary Higgs phase pictures
depending on the scales, $\Lambda_{SU(M)}$ and $\Lambda_{SU(N)}$, at which the
two gauge groups become strong.  If $\Lambda_{SU(N)} \gg \Lambda_{SU(M)}$,
$SU(N)$ behaves like QCD and the symmetry breaking pattern
$SU(M)_{global}\times SU(M)_{gauge} \to SU(M)_{global}$ is expected.  In this
case a $\psi_\alpha \psi_\beta$ dynamical mass forms and the remaining
$\psi_\gamma$ is massless.  Alternatively, if $\Lambda_{SU(M)} \gg
\Lambda_{SU(N)}$, the symmetry breaking $SU(N)_{global}\times SU(N)_{gauge}
\to SU(N)_{global}$ is expected, a $\psi_\beta \psi_\gamma$ dynamical mass
forms, and $\psi_\alpha$ remains massless. In {\it all} cases a massless
$(M,\bar{N})_{+1}$ fermion remains, summarized by the ``reduced moose''
\beq
\includegraphics*[width=2cm]{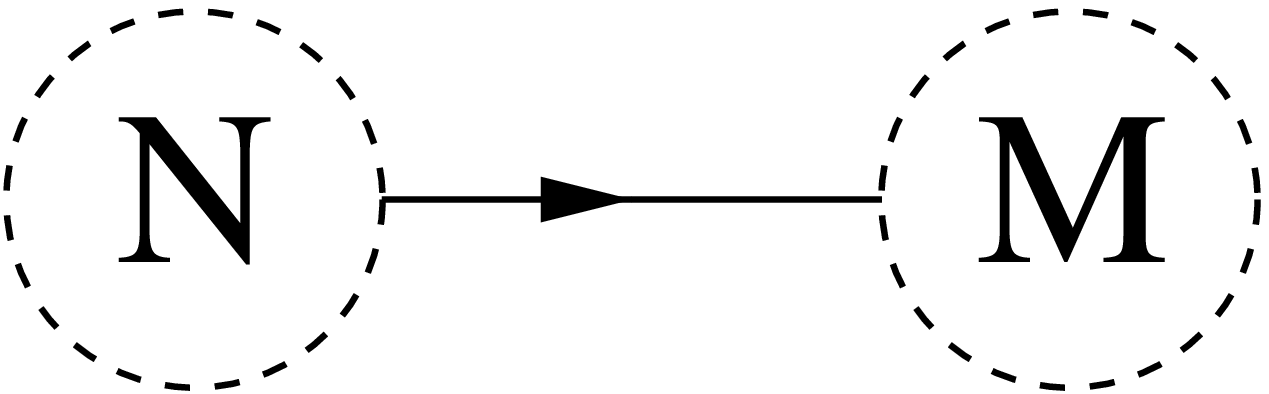}~.
\eeq
Out of these basic ingredients, many
models\cite{Georgi:1986hf,Georgi:1986dw} with composite fermions (and
scalars) can be formed, and the interested reader is encouraged to
explore the literature.

\section{Composite Gauge Bosons: Duality\protect\footnote{For more complete 
    review of duality in supersymmetric theories, see Peskin's
    lectures in the 1996 TASI summer
    school.\protect\cite{Peskin:1997qi}} in SUSY $SU(N_c)$}

Consider a supersymmetric $SU(N_c)$ gauge theory with $N_f$ flavors.
The left-handed fermions $\psi_L$, of an ordinary gauge theory, become
``chiral superfields'' comprised of a complex scalar $Q$ and
left-handed fermion $\psi_Q$. Similarly, the left-handed charge
conjugate fermions of an ordinary gauge theory $\psi^c_L$ become the
fields $\bar{Q}$ and $\psi_{\bar{Q}}$. The gluon $g$ of the ordinary
theory becomes a ``vector superfield'' with the addition of the
gluino, an adjoint Majorana-Weyl fermion $\lambda^a$ ($\lambda_L
\equiv \lambda^c_R$). The global symmetry of the theory is $SU(N_f)_L
\times SU(N_f)_R \times U(1)_B \times { U(1)_R}$, where the fermions
have charges
\beq
\begin{array}{c|c|cccc}
& SU(N_c) & SU(N_f)_L & SU(N_f)_R & U(1)_B & U(1)_R \\
\hline
\psi_Q & N_c & N_f & 1 & 1 & -\,{N_c\over N_f} \\
\psi_{\bar Q} & \bar{N_c} & 1 & \bar{N_f}  & -1 & -\,{N_c\over N_f} \\
\lambda & N^2_c-1 & 1 & 1 & 0 & +1 \\
\end{array}
\eeq
Note that the addition of the massless gluinos has resulted
in an {\it additional} nonanomalous symmetry $U(1)_R$.
The anomaly factor for $U(1)_R SU(N_c)^2$ is
\beq
2\cdot N_f\cdot \left(-{N_c\over N_f}\right)
  {1\over 2} + (+1)N_c \equiv 0
\eeq
where we note that the index of the fundamental is $1/2$ while that of
the adjoint is $N_c$.

Seiberg\cite{Seiberg:1995pq} has conjectured that, for $N_c+1 < N_f <
3 N_c$, this theory is a ``dual'', {\it i.e.} has the same low-energy
theory as, a supersymmetric $SU(N_f-N_c)$ gauge theory with fields
\beq
\begin{array}{c|c|cccc}
& SU(N_f-N_c) & SU(N_f)_L & SU(N_f)_R & U(1)_B & U(1)_R \\
\hline
\psi_q & N_f - N_c & \overline{N_f} & 1 & {N_c \over {N_f-N_c}} & 
{N_c\over N_f}-1 \\

\psi_{\bar{q}} & \overline{N_f - N_c} & 1 & N_f & -{N_c \over {N_f-N_c}} & 
{N_c\over N_f}-1 \\

\psi_M & 0 & N_f & \overline{N_f} & 0 & 1-{2N_c \over N_f} \\

\tilde{\lambda} & (N_f - N_c)^2-1 & 1 & 1 & 0 & +1 \\
\end{array}
\eeq
and a superpotential $W \propto M \psi_q \psi_{\bar{q}}$ coupling the
global symmetries on the dual ``quark'' $q$ and ``meson'' $M$ fields.

All global anomalies match, but only if both the mesons and dual
gauginos are included! Unlike the composite theories we have discussed
previously, the dual gauge bosons and quarks {\it cannot} be
interpreted as simple bound states of fundamental particles.  The
proposed duality satisfies\cite{Peskin:1997qi} a number of other
nontrivial checks as well, including: holomorphic decoupling,
consistency with non-abelian conformal phase $3N_c/2 < N_f < 3 N_c$,
generalization to $N=2$ supersymmetric theories and string
calculations. While no proof has yet been given, the overwhelming
preponderance of evidence indicates that Seiberg duality holds and the
theory described provides a highly nontrivial example of
compositeness.

\bigskip
\noindent\underline{\tt For your consideration \ldots}
\bigskip

Verify that all of the following $SU(N_c)$ anomalies
match in the $SU(N_f-N_c)$ dual theory:

\begin{itemize}

\item $(SU(N_f)_L)^3$

\item $(SU(N_f)_L)^2 U(1)_B$

\item $(SU(N_f)_L)^2 U(1)_R$

\item $(U(1)_B)^2 U(1)_R$

\item ${\rm Tr}\,U(1)_R$ (``gravitational anomaly'')

\item $(U(1)_R)^3$

\end{itemize}

\section{The ACS Conjecture\protect\cite{Appelquist:1999hr}}

Given the panoply of examples we have considered, it is worth asking if there
is a limit on the complexity of the low-energy theory.  Consider the
free-energy per unit volume ${\cal F}$ of a theory at temperature $T$, and
define
\beq
f_{IR} = - \lim_{T\to 0}{{\cal F}\over T^4}{90\over \pi^2}
\eeq
and
\beq
f_{UV} = - \lim_{T\to \infty}{{\cal F}\over T^4}{90\over \pi^2}~.
\eeq
For a free theory, both equal $n_B + {7n_F/8}$.  In this sense, $f_{IR,UV}$
counts the number of low- and high-energy degrees of freedom. Appelquist,
Cohen, and Schmaltz have conjectured\cite{Appelquist:1999hr} that: 
\beq
f_{IR} \le f_{UV}~, 
\eeq
corresponding to the intuitively reasonable result that the number of
degrees of freedom should not {\it increase} at low energies.  This
conjecture has been confirmed for a number of
models,\cite{Appelquist:1999hr,Appelquist:1999vs} but no general proof
has been found.

\section{Conclusions on Compositeness}

\begin{itemize}
  
\item Experimental limits place a lower bound of order 4 TeV (using
  the ELP\cite{Eichten:1983hw} convention) on the {scale of quark and
    lepton compositeness}.
  
\item Weak coupling and asymptotic freedom imply that the standard model
  gauge bosons are likely to be {\it fundamental}; the longitudinal $W$ and
  $Z$ may be composite with a scale of order 1 TeV or higher.
  
\item Composite Scalars: Generically, light scalars occur near a 2nd
  order phase transition. ``Tuning'' is required to keep them light
  compared to the compositeness scale, leading to potential
  hierarchy/naturalness problems!
  
\item Composite Fermions: Massless fermions are a natural consequence
  of confinement and unbroken chiral symmetry. 't Hooft's anomaly
  matching conditions must be satisfied.
  
\item Composite Gauge Bosons: Seiberg duality shows 4-dimensional
  field theory at its most subtle -- who says we need 10 or 11
  dimensions!

\end{itemize}

\bigskip
\noindent{\bf Acknowledgements:}

I thank Jon Rosner and K.~T.~Mahanthappa for organizing a stimulating
summer school, and Gustavo Burdman, Myckola Schwetz, and especially Elizabeth
Simmons for comments on the manuscript.  {\em This work was supported in part
  by the Department of Energy under grant DE-FG02-91ER40676.}

\bibliography{tasi}
\bibliographystyle{h-physrev}

\end{document}